\documentclass[%
 amsmath,amssymb,
 aps,
pra,
]{revtex4-2}
\usepackage{graphicx}
\usepackage{dcolumn}
\usepackage{bm}

\begin{document}

\preprint{APS/123-QED}

\title{I$_4$ complex acting as near-infrared emitter with a quantum memory in silicon}

\author{Miaomiao Han}
\affiliation{School of Science, Huzhou Normal University, Huzhou, 313000 Zhejiang, China\\HUN-REN Wigner Research Center for Physics, Budapest, PoB 49, H-1525 Budapest, Hungary}
\email{mmhan923@mail.ustc.edu.cn}

\author{Song Li}%
\affiliation{Beijing Computational Science Research Center, 100193 Beijing, China}%

\author{Gergő Thiering}
\affiliation{HUN-REN Wigner Research Center for Physics, Budapest, PoB 49, H-1525 Budapest, Hungary}%

\author{Thomas Frauenheim}
\affiliation{Beijing Computational Science Research Center, 100193 Beijing, China\\Contractor University, School of Science, Campus Ring 1, 28759 Bremen, Germany}%

\author{Peter Deák}
\affiliation{Beijing Computational Science Research Center, 100193 Beijing, China\\ HUN-REN Wigner Research Center for Physics, Budapest, PoB 49, H-1525 Budapest, Hungary}
\email{deakbremen@gmail.com}

\author{Bing Huang}
 \affiliation{Beijing Computational Science Research Center, 100193 Beijing, China}%

\author{Adam Gali}
 \affiliation{HUN-REN Wigner Research Center for Physics, Budapest, PoB 49, H-1525 Budapest, Hungary\\Department of Atomic Physics, Budapest University of Technology and Economics,H-1521 Budapest, Hungary\\MTA–WFK Lendület “Momentum” Semiconductor Nanostructures Research Group, P.O. Box 49, H-1525 Budapest, Hungary}%
 \email{gali.adam@wigner.hun-ren.hu}
\date{\today}

\begin{abstract}
Silicon color centers emitting in the near infrared offer a route to spin--photon interfaces compatible with mature complementary metal--oxide--semiconductor fabrication. A key unresolved candidate is the X-line at 1039.7 meV in the photoluminescence (PL) spectrum, long attributed to a complex of four self-interstitials (I$_4$), yet whose microscopic structure and spin properties have remained controversial and uncharacterized at a predictive level. Using hybrid-functional (HSE06) density functional theory, we unambiguously establish the bare I$_4$ complex as the origin of the X-line: our calculations quantitatively reproduce the zero-phonon line, the complete phonon sideband structure---identifying the prominent secondary peaks as one- and two-phonon replicas of strongly coupled quasi-local modes---and the EPR hyperfine constants. The charge transition level is found at VBM$+$0.08~eV, conclusively ruling out the previously proposed association with the B3 EPR center and its correlated deep donor level. Most significantly, we identify a metastable spin-1 triplet excited state separated from the singlet by $\sim$1.6~meV, consistent with the NL51 EPR signal. The $D_{2d}$ symmetry of the defect suppresses intersystem crossing and phosphorescence by first-order spin--orbit coupling, implying an exceptionally long spin--lattice relaxation time. Combined with a spectrally resolvable zero-field splitting and near-infrared emission, the triplet state of the I$_4$ complex constitutes a viable spin quantum memory intrinsic to silicon.  
\end{abstract}

\maketitle


\section{introduction}
Scalable quantum networks require spin--photon interfaces that combine coherent single-spin control with near-infrared photon emission\cite{childs2010quantum,monroe2002quantum,devoret2004implementing,hanson2008coherent,bertaina2007rare}. Point defects in wide-gap semiconductors are natural candidates: the nitrogen-vacancy center in diamond\cite{toyli2013engineering,balasubramanian2009ultralong,gali2019abinitio,pompili2021multinode} and the divacancy in silicon carbide\cite{li2022sic,koehl2011sic,anderson2019control,bian2025theory} have both demonstrated optically addressable spin qubits in proof-of-principle devices. Silicon is emerging as a particularly compelling host: its unmatched fabrication maturity and compatibility with complementary metal--oxide--semiconductor (CMOS) processes offer a natural path to wafer-scale quantum photonic integration\cite{durand2021nir,khoury2022silicon,zhang2020platforms}. Identifying and fully characterizing native silicon color centers is therefore a prerequisite for leveraging this platform in quantum information applications.

Identification of the exact microscopic structure of color centers, as well as in-depth characterization of their magneto-optical properties are the first steps toward their application. The photoluminescence (PL) G-center in silicon has been identified to have a buckled C$_\text{Si}$--Si--C$_\text{Si}$ configuration in the neutral charge state and is a potential qubit in silicon\cite{song1990carbon,davies1989optical,redjem2020telecom,hollenbach2020telecom,baron2022gcenters,hollenbach2022wafer,redjem2023defect,prabhu2023artificial,lefaucher2023cavity,redjem2023naturecomm,durand2024genuine,udvarhelyi2021telecom}. Other carbon-related defects, like the carbon interstitial (C$_\text{i}$)\cite{leary1997carbon,wang2014carbon,deak2024telecom}, the G$^\prime$-center [(C$_\text{Si}$--Si--C$_\text{Si}$)+C$_\text{Si}$]\cite{aberl2024epitaxial,salomon2025led}, the C-center (C$_\text{i}$+O$_\text{i}$)\cite{udvarhelyi2022lband,hayama2004isotope,wen2025arxiv}, or the T-center [(C$_\text{2}$)$_\text{Si}$+H]\cite{ivanov2022localization,bergeron2020interface,higginbottom2022single,safonov1996carbon,dhaliah2022tcenter,xiong2024discovery}, have also been explored. (Here “I” denotes a silicon self-interstitial, the “Si” subscript a substitutional, while the “i” subscript an interstitial impurity.) Two other prominent lines, termed W and X, can be produced by proton- or neutron-irradiation or by silicon-implantation into high purity material, and are assumed to be related to intrinsic self-interstitial aggregates (I$_\text{n}$). The W-center, with strong evidence for an origin from I$_3$, has been thoroughly characterized and is by now well understood\cite{bertaina2007rare,li2022sic,buckley2020wcenters,bao2007led,komza2024indistinguishable,chartrand2018linewidth,tan2003implantation,lee2008interstitial,baron2022wcenters}, but the assignment of the X-line at 1039.7 meV is still controversial\cite{tkachev1977radiative,awadelkarim1990photoluminescence,52hayama2004implantation}.

It has been suggested\cite{arai1997interstitial,54carvalho2005dft,santos2016origin}, that a complex of $D_{2d}$ symmetry, consisting of four silicon self-interstitials (two Si atoms sharing the same lattice site) next to each other, with axes slightly tilted from the [001] direction, might give rise to the X-line due to the recombination of an exciton bound to the neutral defect. This, so-called I$_4$ complex (Fig. 1), in its positive charge state, has been proposed to be the origin of an EPR (electron paramagnetic resonance) signal, usually called B3\cite{pierreux2005esr,brower1976epr,mchedlidze2003tetra}. Calculations on I$_4$ could reproduce some local vibrational modes (LVMs), observed as phonon replicas in the fine spectrum of the X-line\cite{52hayama2004implantation} (but also in infrared absorption\cite{59newman1975optical}, see Table \ref{table2}) and supported this claim. However, the EPR signal B3 has shown correlation with a deep donor charge transition level (CTL), observed by  DLTS (Deep Level Transient Spectroscopy) at 0.29 eV above the VBM (Valence Band Maximum)\cite{54carvalho2005dft,mukashev1982defects,coomer2001tetra}. The obvious controversy between a (+/0) CTL at 0.29 eV above the VBM, and the bound-exciton recombination energy of 1.04 eV (the sum being much larger than the gap of silicon), was noted but not resolved.

\begin{figure}[htbp]
\centering
\includegraphics[width=0.3\linewidth]{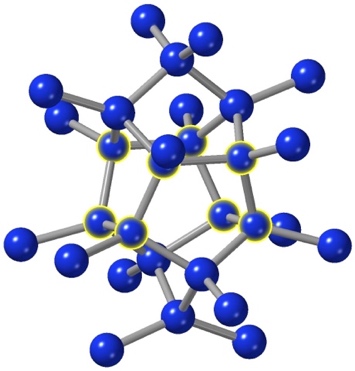}
\caption{Structure of the I$_4$ complex. The four split-interstitials are highlighted.}
\end{figure}

In a more recent paper\cite{kiyoi2022annealing}, the correlation of the X-line with the shallow hydrogen donors\cite{myers1992hydrogen} (and refs. therein) has been reported, and the center was reassigned to a hydrogenated I$_4$ complex. The measured EPR signal was similar to, but slightly different from, the B3 center\cite{kiyoi2023hydrogen}. Obviously, the question then arises what kind of a PL the bare I$_4$ defect would show. It should also be checked whether it has a triplet excited state, similar to the case of the C$_i$ defect, the zero-field-splitting (ZFS) of which can be used as spin memory\cite{deak2024telecom}.

To our knowledge, theoretical works which investigated the I$_4$ defect\cite{wen2025arxiv,chartrand2018linewidth,tan2003implantation,lee2008interstitial,santos2016origin} so far never went beyond the local- or semi-local approximation of the exchange functional in density functional theory (DFT). Due to the underestimation of the band gap and of the localization of the defect wave function, quantitatively reliable charge transition levels, PL energies, and hyperfine parameters could not be obtained, preventing a definitive assignment and any assessment of the spin properties relevant for applications. In this paper we report hybrid-functional (HSE06) calculations that settle the assignment of the X-line to the bare I$_4$ complex, resolve the long-standing contradictions with EPR and DLTS data, and—most importantly for applications—characterize the metastable triplet excited state as a candidate quantum memory in silicon. In Sec.~II we describe the theoretical methods; in Sec.~III we present and discuss our results on the optical properties, the charge transition level, the EPR signal, and the spin properties of I$_4$; Sec.~IV summarizes our conclusions.

\section{methods}
As was shown earlier\cite{65deak2010hse}, the HSE hybrid functional with the original parametrization (HSE06) satisfies in silicon the two conditions required to calculate defect properties accurately: it reproduces the band gap and is Koopmans-compliant\cite{66deak2019hybrid}. We have carried out such HSE(0.25,0.20) calculations with the Vienna ab-initio simulation package (VASP), using the projector augmented wave method\cite{67kresse1994md,68kresse1996vasp,69kresse1999paw}, and a plane-wave cutoff of 420 eV. Defects were modelled in a 512-atom supercell (4$\times$4$\times$4 of the conventional Bravais cell), and all atoms were allowed to relax in a constant volume till the forces were below 0.01 eV/Å. The $\Gamma$-point approximation was used for Brillouin-zone sampling. The equilibrium lattice constant and the 0~K gap were found  to be 5.4333 \AA ~and 1.16 eV, respectively (in good agreement with experiment). The total energy of the charged defect has been calculated by applying the SCPC method for charge correction\cite{70chagas2021correction}. Charge correction was also used in the case of the bound exciton (where the electron wave function is delocalized), as described in Sec. 3.3.4 of Ref.\cite{71gali2023review}. The zero-phonon line (ZPL) was obtained as the energy difference of the relaxed ground and the relaxed excited state, the latter calculated with the constrained occupation (cDFT) or $\Delta$SCF method. The excited state created this way is not a spin eigenstate, so to obtain the correct ZPL for the singlet-to-singlet transition, the correction derived in Ref. \cite{72mackoit2019carbon} (Supplementary Material) was applied. In the case of the bound exciton, a band-filling correction was also used\cite{73lany2008assessment}, since the $\Gamma$-point in the reduced Brillouin zone of the supercell represents the [0.75,0,0] primitive k-point explicitly, while the real CBM is at [0.83,0,0]. Furthermore, the singularity in the Coulomb kernel for exact-exchange is circumvented in VASP by an expansion with auxiliary functions\cite{74massidda1993hf},but using the $\Gamma$-point alone causes an artificial deepening of conduction-band-derived occupied states\cite{75broqvist2009erratum}. Therefore, the total energy of the excited state was corrected with the energy difference of this level and its unoccupied counterpart\cite{76ho2018ga2o3}.

The spectrum of the phonon replicas was computed by the generating function method\cite{77alkauskas2014lineshape} at $T=30$ K, based on vibration calculations using the Perdew--Burke--Ernzerhof (PBE) functional\cite{78perdew1996gga}. The radiative lifetime was calculated as the inverse of
\begin{equation}
\label{eq:gamma_rad}
\Gamma_{\mathrm{rad}} = \frac{n_D E_{\mathrm{ZPL}}^3 {\mu}^2}{3\pi\epsilon_0c^3{\hbar}^4}
\end{equation}
(see Ref.~\cite{79li2022ws2}), where the optical transition dipole moment $\mu$ was calculated by taking the respective Kohn-Sham wave functions representing the electronic ground and excited states, and $n_D=3.485$ is the refractive index of silicon at 1.45 $\mu$m or $\approx0.85$ eV (see Ref. \cite{80green2008silicon}). $E_\text{ZPL}$ is the observed energy of the ZPL, $\epsilon_0$ is the dielectric permittivity of vacuum, $c$ is the speed of light and $\hbar$ is the reduced Planck constant. The methods of calculating hyperfine tensors and the zero-field-splitting tensor\cite{81bodrog2013zfs,82szasz2013hyperfine}, were implemented into VASP by M. Marsman.

The spin--orbit coupling (SOC) matrix elements connecting the relevant singlet and triplet many-electron states were obtained from non-collinear DFT calculations including spin--orbit interaction, by reducing the two-particle matrix elements to single-particle matrix elements between the corresponding Kohn--Sham orbitals (see Sec.~III). The intersystem crossing (ISC) rate was then estimated within the framework of Fermi's golden rule at $T=0$ K. We adopt the golden-rule expression with the $2\pi/\hbar$ prefactor, 
\begin{equation}
\label{eq:gamma_isc}
\Gamma_{\mathrm{ISC}} = \frac{2\pi}{\hbar}\,|H_{\mathrm{SOC}}|^{2}\,F,
\end{equation}
where $\hbar$ is the reduced Planck constant, $H_{\mathrm{SOC}}$ is the symmetry-allowed (perpendicular) SOC spin–orbit coupling matrix element between the singlet and the $m_s=\pm1$ triplet sublevels, and $F$ is the Franck--Condon weighted density evaluated at the relevant electronic energy gap $\Delta E$. The Franck–Condon weighted density of states is obtained from the Fourier transform of the generating function
\begin{equation}
\label{eq:gamma_fcdw}
F(E) = \frac{1}{2 \pi \hbar} \int{dt\ e^{iEt/\hbar} G(t)} 
\end{equation}
\begin{equation}
\label{eq:gamma_Gt}
G(t) = \exp\left[\sum_i S_i(e^{-i\omega_it}-1)\right],
\end{equation}
where $S_i$ denotes the partial Huang--Rhys factor of the $i_{th}$ vibrational mode and $\omega_i$ is the phonon frequency taken from the ground-state vibrational analysis. The mass-weighted displacement vector between the singlet and triplet geometries was projected onto the common vibrational basis to get $S_i$, with the relaxed geometry of the $^1B_2$ singlet serving as a proxy for that of the $^1E$ state, as justified in Sec.~\ref{sec:qmemory}.
With this definition $\Gamma_{\mathrm{ISC}}$ is the angular rate (the inverse lifetime), which we equivalently write as  $\Gamma_{\mathrm{ISC}}=2\pi X$, where $X=\Gamma_{\mathrm{ISC}}/2\pi$ is the corresponding ordinary-frequency value quoted in Sec.~III. This procedure yields an order-of-magnitude estimate of the ISC rate, which is sufficient to assess the feasibility of triplet population.
 
 \section{RESULTS AND DISCUSSION}
 \subsection{Electronic structure of the ground and excited states}
\label{sec:elstruct}
\begin{figure}[htbp]
\centering
\includegraphics[width=0.85\linewidth]{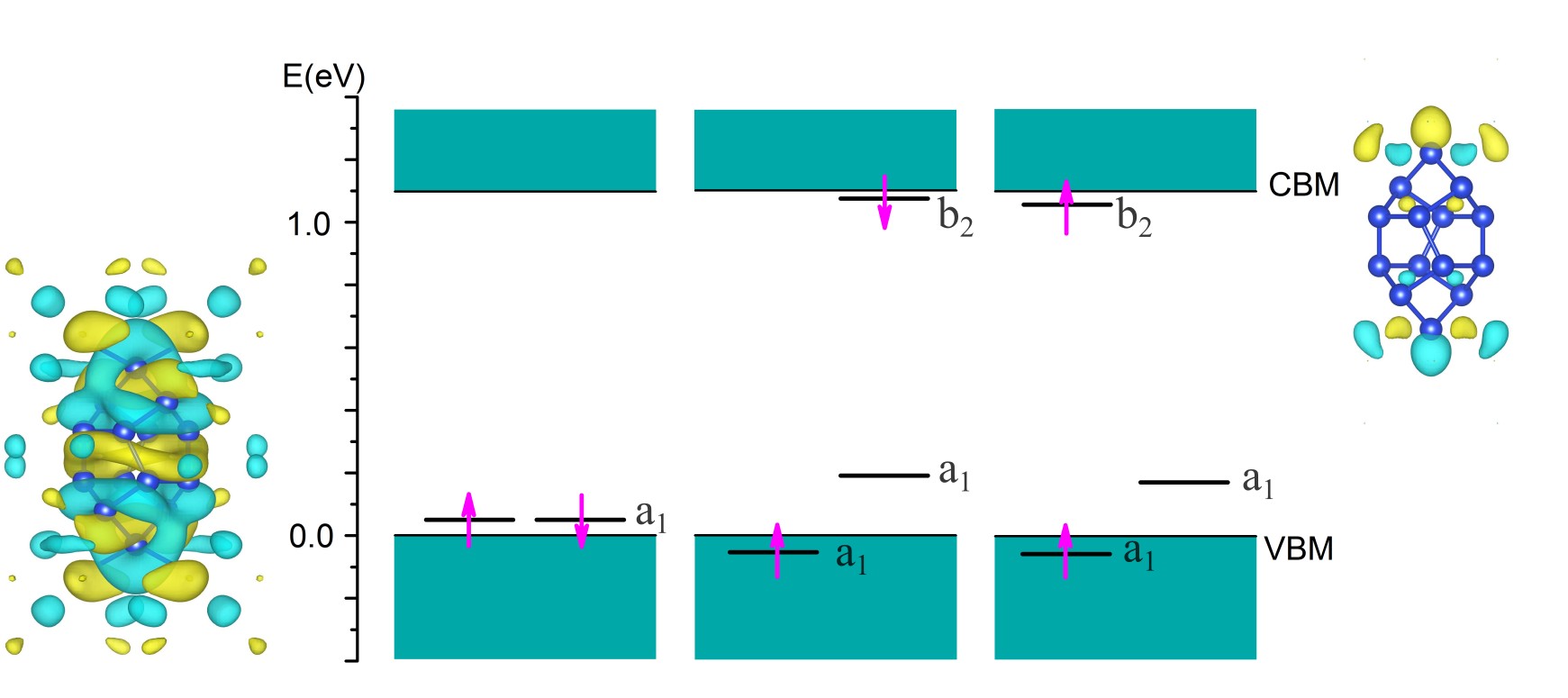}
\caption{Single-particle level structure of the I$_4^0$ defect in the spin-polarized HSE06 calculation, showing the ground state ($^1A_1$), singlet ($^1B_2$), and triplet ($^3B_2$) excited states; the left and right insets show the wavefunctions of the $a_1$ and $b_2$ levels (isosurface value $2.8\times10^{-8}$).}
\label{fig2}
\end{figure}

In the ground state of the neutral defect I$_4^0$, we find a doubly occupied electron state, localized to I$_4$ with a similar distribution of the wave function as reported earlier\cite{coomer2001tetra}, and with a level above the VBM, as shown in Fig.~\ref{fig2}. The association of a donor level with this fully-coordinated structure was explained as a result of the compressive strain in the defect core. Bond compression enhances the overlap between local $sp^3$ orbitals, modifying the local bonding--antibonding splitting and causing some states to split off from the bulk bands into the band-gap region.

In pristine silicon, the CBM is not at $\Gamma$, but lies along the $\Delta$ line (as mentioned in Methods). The little group along $\Delta$ is smaller. For the supercell, folded $\Delta$ states appear at $\Gamma$, but they form representations induced from the original little group at $\Delta$ and valley degeneracy breaks. With a $D_{2d}$ defect, the valley splits into $2a_1\oplus b_1\oplus b_2\oplus e$. The ground state with two electrons occupying the $a_1$ level belongs to the irreducible representation $^1A_1$. Promoting one electron from the $a_1$ level into any of these CBM-derived valley-orbit states produces an excited configuration, each having both a singlet and a triplet counterpart. We focus first on the lowest-energy excited state, in which the electron is promoted to the $b_2$ level---the lowest valley-orbit state, arising from the in-plane antisymmetric combination---whose singlet and triplet transform as $^1B_2$ and $^3B_2$, respectively. The remaining valley-orbit states ($2a_1$, $b_1$, and $e$) give rise to analogous singlet--triplet manifolds ($^{1,3}A_1$, $^{1,3}B_1$, and $^{1,3}E$) at higher energy. The irreducible representations of the lowest states, formulated in the two-electron representation in the orbital and spin product basis, are as follows. Throughout, $a$ and $b$ denote the $a_1$ and $b_2$ single-particle orbitals; the $m_s = \pm 1$ triplet sublevels are written in the real (Cartesian) basis $\lvert\uparrow\uparrow \rangle \pm \lvert \downarrow \downarrow\rangle$, which is the natural basis for the zero-field-splitting and spin--orbit analysis below.
\begin{equation}
\label{eq:1A1}
\lvert^1A_1\rangle=\lvert {aa} \rangle \otimes \frac{1}{\sqrt{2}} (\lvert\uparrow\downarrow\rangle - \lvert\downarrow \uparrow\rangle)
\end{equation}
\begin{equation}
\label{eq:1B2}
\lvert^1B_2\rangle=\frac{1}{\sqrt{2}}(\lvert{ab} \rangle +\lvert {ba} \rangle ) \otimes \frac{1}{\sqrt{2}}(\lvert\uparrow\downarrow\rangle - \lvert\downarrow \uparrow\rangle)
\end{equation}
\begin{equation}
\label{eq:3B2_0}
\lvert^3B_2, m_s=0\rangle=\frac{1}{\sqrt{2}}(\lvert{ab} \rangle -\lvert {ba} \rangle ) \otimes \frac{1}{\sqrt{2}}(\lvert\uparrow\downarrow\rangle + \lvert\downarrow \uparrow\rangle)
\end{equation}
\begin{equation}
\label{eq:3B2_pm1}
\lvert^3B_2,m_s=\pm1 \rangle=\frac{1}{\sqrt{2}}(\lvert{ab} \rangle -\lvert {ba} \rangle ) \otimes \frac{1}{\sqrt{2}}(\lvert\uparrow\uparrow \rangle \pm \lvert \downarrow \downarrow\rangle)
\end{equation}
Among the higher-lying excited states, the singlet derived from the $e$ valley-orbit level (with $e$ denoting one of its components) plays a special role in the intersystem crossing discussed in Sec.~\ref{sec:qmemory}, and is written analogously as
\begin{equation}
\label{eq:1E}
\lvert^1E\rangle=\frac{1}{\sqrt{2}}(\lvert {ae}\rangle + \lvert{ea}\rangle) \otimes \frac{1}{\sqrt{2}}(\lvert\uparrow\downarrow \rangle- \lvert\downarrow \uparrow\rangle)
\end{equation}

\subsection{Fluorescence spectrum and identification}
We have calculated the recombination energy of a bound exciton, with the hole localized on I$_4$ and the electron being on the conduction band minimum (CBM). As it was recently shown that the $\Delta$SCF method in VASP is sensitive to the starting wave function when relaxing the excited state, and that using a PBE result for that is preferable\cite{83xiong2025dscf}, we adopted the PBE wave function of the hole at the HSE geometry of I$_4^+$ as the starting point. The calculated ZPL and the applied corrections are shown in Table~\ref{table1}.

\begin{table*}[hbtp]
\centering
\caption{\label{table1}%
Calculated ZPL using the PBE starting wave function for relaxing the excited state, and the applied corrections. All values in meV.}
\begin{ruledtabular}
\begin{tabular}{cccccc}
ZPL$_{\mathrm{raw}}$ & \multicolumn{4}{c}{Correction} & ZPL$_{\mathrm{corrected}}$\\
\hline
       & Charge & Spin state & Band filling & Singularity & \\
 980.4 & 57.5   & 11.8       & -47.3        & 31.7        &1034.1\\
\end{tabular}
\end{ruledtabular}  
\end{table*}

Indeed, the ZPL obtained is close to the experimentally observed value of the X-line, 1039.7 meV, but the magnitude of the applied corrections introduces some uncertainty. Therefore, we have calculated the LVMs of the ground state of  I$_4$ and, based on that, simulated the PL spectrum next to the X-line using the PyPhotonics package\cite{84tawfik2022pyphotonics}. 

Table \ref{table2} shows that our calculation for the I$_4$ complex results in similar values for LVMs above the Raman frequency to those published in the literature\cite{54carvalho2005dft}, but somewhat closer to the infrared (IR) experimental data\cite{59newman1975optical}.  We identified 11 local modes with frequencies above the Raman frequency (523  cm$^{-1}$) between 527.2 and 548.7 cm$^{-1}$. These modes are mainly the vibrations of the four split self-interstitial atoms (see l.h.s. of Fig.~\ref{fig:pl}), among which two modes of $A_1$ symmetry are at 548.7 cm$^{-1}$ and 538.6 cm$^{-1}$, and two $E$-modes at 541.4 and 548.0 cm$^{-1}$. The calculated $^{30}$Si isotope shifts of all these modes agree very well with the experimental values. In addition, we find some low-frequency quasi-local $A_1$ vibrational modes, involving both the defect core and nearby lattice atoms.

\begin{table*}[htbp]
\centering
\caption{\label{table2}%
The calculated LVMs and their downward $^{30}$Si isotope shifts (in brackets) , compared to the infrared active modes observed in absorption and as phonon replicas of the ZPL in PL. The mass number of  $^{30}$Si atoms is taken to be 29.9737 a.u.  All frequencies are in cm$^{-1}$.}
\begin{ruledtabular} 
\begin{tabular}{cccccc}
Symmetry & Carvalho et al.\cite{54carvalho2005dft} & present & exptl. IR\cite{59newman1975optical}
 & exptl. PL \cite{52hayama2004implantation} \\
\hline
A$_1$ & 559.4 (18.5)  & 548.7  (17.6)  &  &     556.5 (S$_3$)  (16.9) \\
A$_1$ & 545.8  (18.1) &	538.6  (17.2)  &  &     533.9 (S$_1$)  (16.9) \\
E     & 547.1  (18.1) &	548.0  (17.5)  &  550 &	547.7 (S$_2$)  (18.6) \\
E	  & 546.1  (18.1) & 541.4  (17.3)  &  530 & \\
\end{tabular}
\end{ruledtabular}  
\end{table*}

\begin{figure}[htbp]
\centering
\includegraphics[width=0.55\linewidth]{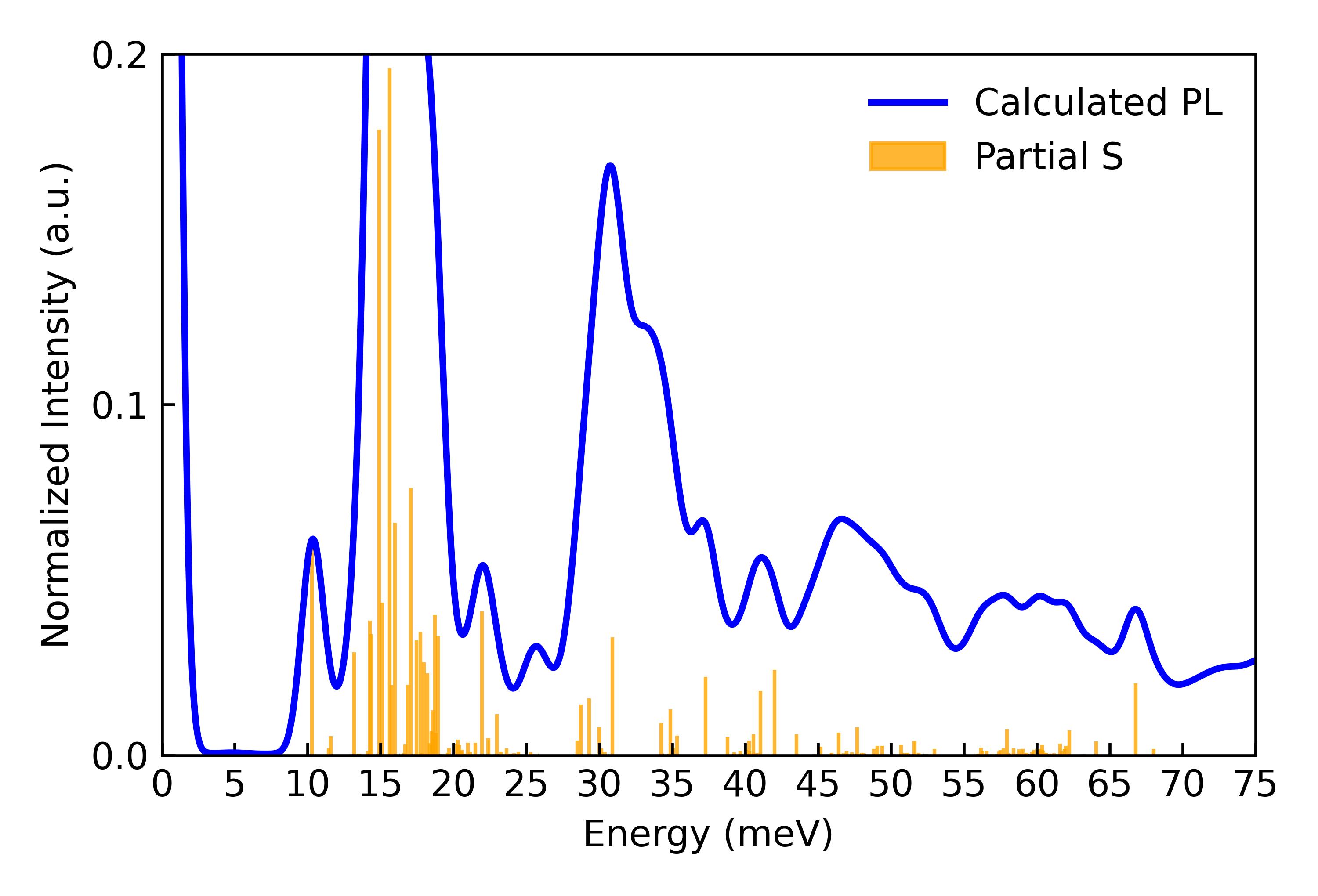}
\caption{Calculated PL sideband relative to the ZPL energy and the Si partial Huang--Rhys factors.}
\label{fig:hr}
\end{figure}

\begin{figure*}[htbp]
\centering
\includegraphics[width=0.9\linewidth]{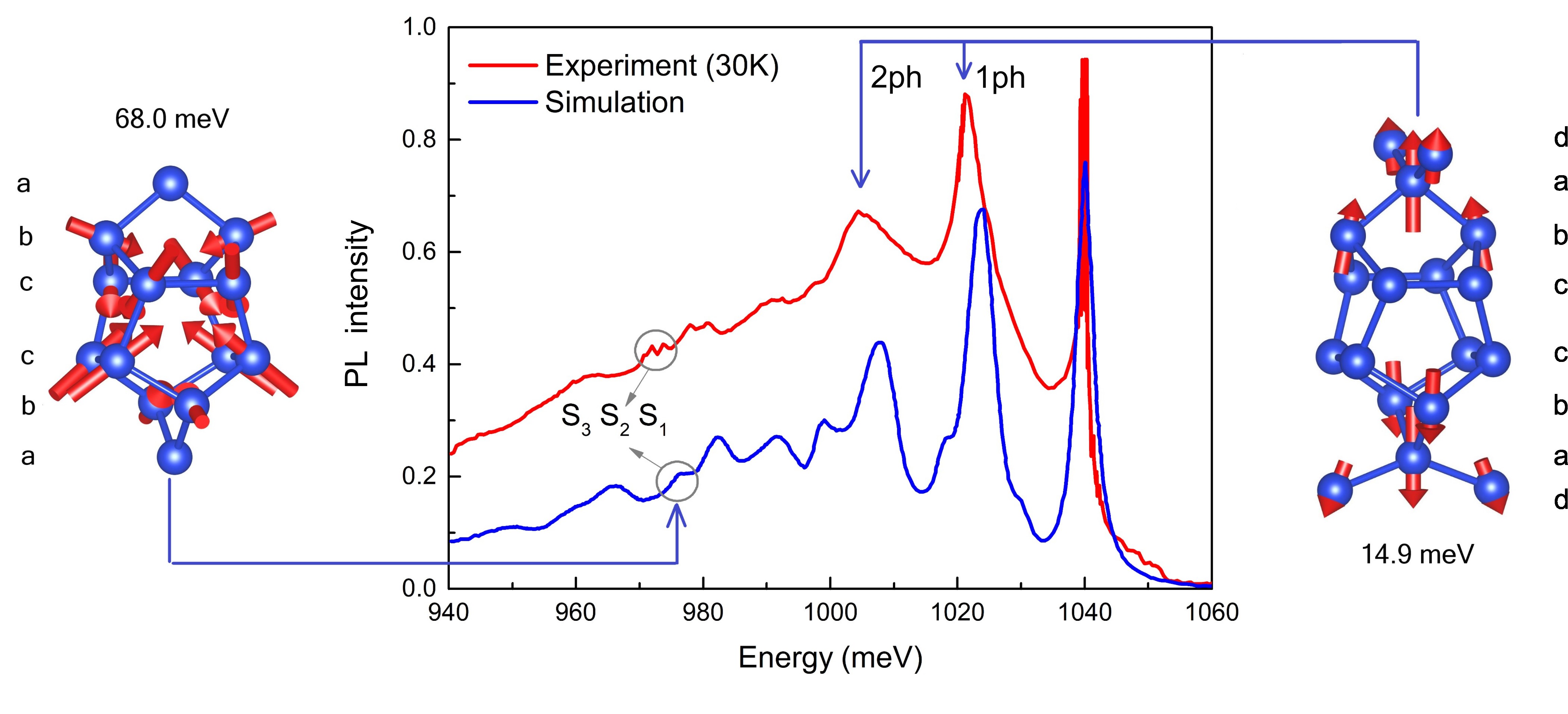}
\caption{The measured (red curve; recorded at 30 K on proton-irradiated natural-isotope silicon, digitized from Ref.~\onlinecite{52hayama2004implantation}) and simulated (blue curve; $T=30$ K) fine structure of the X-line. The LVMs responsible for the phonon replicas of the ZPL are shown schematically.}
\label{fig:pl}
\end{figure*}

The electron--phonon coupling that shapes the sideband is quantified by the partial Huang--Rhys factors shown in Fig.~\ref{fig:hr}. Although the two high-frequency $A_1$ modes are strongly localized, their Huang--Rhys factors remain small, giving weak intensities in the PL spectrum. By contrast, the $E$ modes are symmetry-forbidden within the high-symmetry $D_{2d}$ configuration and exhibit negligible electron--phonon coupling, demonstrating that strong phonon localization alone is not sufficient to ensure strong electron--phonon coupling. The dominant contribution to the Huang--Rhys factor ($S\approx1.42$) instead comes from the low-frequency quasi-local $A_1$ vibrational modes in the range 14.9--18.7 meV (151.1--120.1 cm$^{-1}$), where the significant geometric displacement between the ground and excited states produces strong coupling.

The simulated PL spectrum that results from these couplings is compared to the experimentally observed one in Fig.~\ref{fig:pl}. As can be seen, the agreement is strikingly good. Both curves in Fig.~\ref{fig:pl} correspond to $T=30$ K, the temperature at which the experimental spectrum was recorded. The thermal occupation of the coupled modes (all at $\geq$14.9 meV) stays below $3\times10^{-3}$ at this temperature, so that phonon absorption contributes less than 0.5\% of the sideband intensity and the computed lineshape is essentially its $T=0$ limit; it remains quantitatively valid up to $\approx$50 K, above which anti-Stokes replicas and a reduced Debye--Waller factor would become appreciable. Experimentally, three small phonon replicas (S$_{1,2,3}$) have been resolved\cite{52hayama2004implantation}, while the calculated PL spectrum exhibits only a single small peak, originating from two totally symmetric $A_1$ LVMs that are close in energy at 68.0 and 66.8 meV. It has been speculated that the third phonon replica observed experimentally is caused by the anharmonic coupling of the two highest-energy $E$ modes\cite{54carvalho2005dft}. In addition, local symmetry breaking and/or Herzberg--Teller vibronic coupling may dynamically relax selection rules, thereby activating these $E$ modes and giving rise to the observed phonon sidebands. The two $A_1$ LVMs at 66.8 and 68.0 meV, together with the $E$ mode at 67.9 meV activated via Herzberg--Teller coupling, thus provide a natural explanation for the three observed phonon replicas S$_{1,2,3}$.

A still outstanding issue is the origin of the two pronounced peaks beyond the ZPL, which were not discussed before in the experimental paper. We note that the second peak has a similar lineshape and approximately twice the energy separation from the ZPL as the first peak, suggesting a higher-order phonon replica. The energy spacing of the first peak closely matches that of the low-frequency quasi-local $A_1$ vibrational modes that dominate the Huang--Rhys factor (Fig.~\ref{fig:hr}). Accordingly, the first peak can be attributed to the first (single-phonon) replica of these low-frequency quasi-local $A_1$ modes, while the second peak corresponds to their second phonon replica.

Our results unambiguously prove that the origin of the X-line is the bare I$_4$ complex.

\subsection{Positive charge state: relation to the DLTS and B3 EPR centers}
\label{sec:dlts_b3}

The defect may also exist in its positive charge state, obtained by removing an electron from the $a_1$ level. Ionizing this level does not break the $D_{2d}$ symmetry, but spin polarization splits it: the occupied state goes below the VBM, while the unoccupied one moves higher. (Without spin polarization the singly occupied level is slightly above the VBM.) The probability distribution of the singly occupied state is qualitatively similar to that of the doubly occupied state of the neutral defect. The calculated charge transition level is $\epsilon$(+/0) = VBM + 0.08 eV, consistent with the bound-exciton nature of the X-line; the same value was obtained in a larger 1000-atom supercell (with a lower energy cutoff of 260 eV), confirming its robustness.

The deep donor level at VBM + 0.29 eV observed by DLTS, and the B3 EPR center, were both previously associated with I$_4$ in its positive charge state. Our charge transition level lies far below the DLTS level at VBM + 0.29 eV that anneals out together with the B3 center; I$_4$ can therefore not give rise to that DLTS level. To test the B3 assignment directly, we calculated the hyperfine tensor of I$_4^+$. Table \ref{table3} shows values averaged over the two perpendicular-to-main-axis directions (since EPR observes the average value of the atoms in the same shell of $D_{2d}$ symmetry). Agreement with the HD2\cite{kiyoi2023hydrogen} center can be excluded at once. The $A_\parallel$ values are comparable to the B3 center\cite{pierreux2005esr,brower1976epr,mchedlidze2003tetra} but the $A_\perp$ values are substantially smaller. In addition, the calculation predicts four detectable shells of atoms, while only three have been observed for the B3 center. Therefore, we cannot confirm I$_4^+$ as the origin of the observed EPR signatures.

\begin{table}[htbp]
\centering
\caption{\label{table3}%
The calculated hyperfine tensor based on the I$_4^+$ wave functions, in comparison with experiments.  All values in MHz. The silicon atoms in the core of the defect are classified, from top to bottom of the insert in Fig.~\ref{fig:pl}, as $d$, $a$, $b$, $c$, $c$, $b$, $a$, and $d$.}
\begin{ruledtabular} 
\begin{tabular}{cccccccc}
  & \multicolumn{2}{c}{VASP}   & \multicolumn{2}{c}{B3~\cite{pierreux2005esr,brower1976epr,mchedlidze2003tetra}} & \multicolumn{2}{c}{HD2\cite{kiyoi2023hydrogen}} \\
\hline
$^{29}$Si & $A_\parallel$ & $A_\perp$  & $|A_\parallel|$ & $|A_\perp|$ & $|A_\parallel|$ & $|A_\perp|$ \\
$a$	&   $-36.0$ & 1.1 &   38.4 & 21.3 & 26.3 &11.2 \\
$b$ &	14.6    & 3.9 &   15.2 & 14.0 & 9.0  & 5.6 \\
$c$	&   $-13.8$ & 2.1 &   8.2  & 6.2  &     &  \\
$d$ &   $-13.0$ & 2.3 &   &      &      &     & \\
\end{tabular}
\end{ruledtabular}  
\end{table}

\subsection{Lowest-energy triplet state as a quantum memory}
\label{sec:qmemory}

\subsubsection{Identification of the triplet state}
\label{sec:ident}

Besides Si-B3, another EPR signal, called Si-NL51, was observed upon illumination and was suggested to arise from the $S=1$ excited neutral state of I$_4$\cite{85stallinga1994epr,mchedlidze2003tetra}. The exchange splitting energy between the excited singlet and triplet states ($\Delta E_{\mathrm{ST}} = E_S - E_T$, with the triplet lying below the singlet) was experimentally fitted to be 1.58 meV from the temperature dependence of the NL51 intensity\cite{mchedlidze2003physb}. We also find a triplet excited state of I$_4$, which is 0.95 eV (uncorrected) above the ground state. The computed splitting exhibits strong dependence on the supercell size and decreases with increasing cell size, reflecting reduced wavefunction overlap as the excited-state wavefunctions become progressively delocalized in larger periodic cells (125.1 meV, 13.9 meV, and 7.3 meV in 216-, 512-, and 1000-atom cells, respectively). The 216-atom supercell is too small to accommodate the extended strain field of the I$_4$ complex without significant periodic-image artifacts and is therefore excluded from the finite-size scaling analysis. For the two larger supercells, the size dependence is well approximated by an effective exponential form $\Delta E(L)=\Delta_{\infty}+Ae^{-L/{\xi}}$, where $L$ is the supercell length and $\xi$ is the characteristic localization length of the excited-state wavefunctions. Taking the experimental asymptote $\Delta_{\infty}=1.58$ meV and fitting the 512- and 1000-atom data to this form yields a localization length $\xi\approx 7.1$ \AA{}, confirming that the computed size dependence is fully consistent with the experimental singlet--triplet gap. This localization length implies that supercells of approximately $9^3$--$10^3$ unit cells ($\sim$49--54 \AA{} in diameter) are required to achieve singlet--triplet splittings within $\sim$0.2--0.1 meV of the converged value.

The zero-field splitting (ZFS) of this triplet provides a further fingerprint for its identification. The calculated ZFS tensor has the parameters $D = 22.25$ MHz and $E=0$, in order-of-magnitude agreement with the experimentally reported $D$ value of NL51 ($\sim$57 MHz), which allows the $m_s$ sublevels to be spectroscopically resolved\cite{85stallinga1994epr}. The physically small value of $D$ originates from the character of the bound-exciton triplet: the hole is localized on the I$_4$ core (Frenkel-like), while the electron occupies the delocalized CBM state (Wannier-Mott-like). This combination produces a large effective spin--spin separation that strongly suppresses the dipolar coupling ($D\propto\langle r^{-3}\rangle$), as also reflected in the extended wavefunction of the $b_2$ level (Fig.~\ref{fig2}, right). Exact numerical agreement with the experimental $D$ is not expected: the two-particle spin-density matrix entering the dipolar spin--spin coupling can only be constructed approximately from the Kohn--Sham orbitals, and an accurate description of the bound-exciton wavefunctions would require supercells in the $\sim$49--54 \AA{} range derived above for the singlet--triplet splitting, which is computationally prohibitive at the hybrid-functional level. Since $E=0$, the $m_s=\pm 1$ levels are degenerate, and an external magnetic field along the $z$ axis is needed to split them.

For future experimental characterization, we also calculated the hyperfine tensor of the triplet, listed in Table~\ref{table4}. We note that no hyperfine structure of NL51 was resolved experimentally\cite{mchedlidze2003tetra}. This is consistent with our results: owing to the severe inhomogeneous linewidth broadening, the resonance profile of this triplet forms a broad envelope without resolved structure, whereas the computed maximum hyperfine splitting ($\sim$1.0 mT, $A_\parallel\approx28.7$ MHz) is small enough that the discrete $^{29}$Si satellite lines are convoluted into the broad envelope and remain unresolved.

\begin{table}[htbp]
\centering
\caption{\label{table4}%
The calculated hyperfine tensor of the triplet excited state of the I$_4$ complex. All values in MHz. The silicon atoms in the core of the defect are classified, from top to bottom of the insert in Fig.~\ref{fig:pl}, as $d$, $a$, $b$, $c$, $c$, $b$, $a$, and $d$.}
\begin{ruledtabular} 
\begin{tabular}{cccccc}
$^{29}$Si	 & $A_\parallel$	& $A_\perp$ \\
\hline
$a$	&   $-28.7$ &	$-9.0$ \\
$b$ &	5.5     & $-0.3$ \\
$c$	&   $-7.0$  & 1.2 \\
$d$ &  $-6.8$   & 1.5 
\end{tabular}
\end{ruledtabular}  
\end{table}

\subsubsection{Population and decay mechanism}
\label{sec:mech}

\begin{figure}[htbp]
\centering
\includegraphics[width=0.6\linewidth]{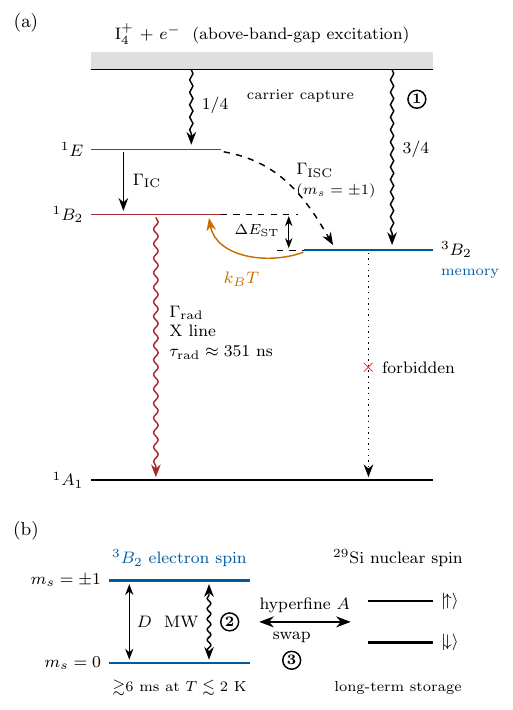}
\caption{(a) Population and decay pathways of the bound exciton of I$_4$ (energies not to scale). After above-band-gap excitation, sequential capture of carriers with uncorrelated spins populates the triplet and singlet manifolds in a 3:1 ratio (step 1). Within the singlet branch, the spin-conserving internal conversion ($\Gamma_{\mathrm{IC}}$) from $^1E$ to $^1B_2$ competes with the spin--orbit-driven intersystem crossing ($\Gamma_{\mathrm{ISC}}$) from $^1E$ to the $m_s=\pm1$ sublevels of the metastable triplet $^3B_2$; the singlet $^1B_2$ decays radiatively at rate $\Gamma_{\mathrm{rad}}$ (X-line). The phosphorescence $^3B_2\rightarrow{}^1A_1$ is symmetry-forbidden, so that the population stored in the triplet is lost only by thermal activation across the singlet--triplet gap $\Delta E_{\mathrm{ST}}$ (orange arrow). (b) Operation of the quantum memory within the $^3B_2$ manifold: the sublevels split by the zero-field splitting $D$ are initialized and manipulated by microwaves (MW, step 2), and the hyperfine interaction $A$ with a weakly coupled $^{29}$Si nuclear spin allows the stored state to be swapped onto a long-lived nuclear ancilla (step 3).}
\label{fig:mech}
\end{figure}

Having established the triplet as a metastable, spectroscopically identifiable state, we now consider how it is populated and how it decays. The pathways are summarized in Fig.~\ref{fig:mech}(a). In this subsection the mechanism is developed on symmetry and kinetic grounds only; the numerical values of the rates and the assessment of their reliability are deferred to Sec.~\ref{sec:rates}. Under above-band-gap excitation the bound exciton is formed by sequential carrier capture at the defect and then relaxes through the valley--orbit ladder of excited states introduced in Sec.~\ref{sec:elstruct}. Population can reach the $^3B_2$ triplet in two ways: directly, through the spin statistics of carrier capture, or from the singlet branch by intersystem crossing (ISC).

The first route follows from spin statistics. The electron captured by I$_4^+$ is uncorrelated with the spin of the bound hole, so the bound exciton forms in the triplet manifold in 3/4 and in the singlet manifold in 1/4 of the capture events, provided that the capture rates into the individual sublevels are equal (an assumption examined in Sec.~\ref{sec:rates}). The triplet branch relaxes by spin-conserving internal conversion (IC) within the triplet valley--orbit ladder down to the metastable $^3B_2$ state, whereas the singlet branch relaxes toward $^1B_2$ and decays radiatively, feeding the X-line emission.

The second route proceeds through the singlet branch and is governed by spin--orbit coupling (SOC). Direct ISC from the singlet bound exciton $^1B_2$ to the triplet $^3B_2$ is symmetry-forbidden by first-order SOC, as shown below; the singlet branch can therefore feed the triplet only through a higher-lying singlet state. The $^1E$ state, derived from the $e$ valley-orbit level, is transiently populated during the capture cascade and can play this role. This is favorable on symmetry grounds: by El-Sayed's rule\cite{elsayed1963soc}, ISC between states of the same orbital character (e.g.\ $^1E\to{}^3E$) is strongly suppressed, whereas the $^1E\to{}^3B_2$ channel changes the orbital character and is comparatively efficient. Spin--orbit coupling is described by the Hamiltonian operator
\begin{equation}
\label{eq:Hso}
{\hat{H}}_{\mathrm{so}}=\lambda_z{{\hat{L}}_z}{{\hat{S}}_z}+\lambda_+{{\hat{L}}_+}{{\hat{S}}_-}+\lambda_-{{\hat{L}}_-}{{\hat{S}}_+},
\end{equation}
which belongs to the totally symmetric $A_1$ irreducible representation in the double group. The symmetry selection rules act on the orbital factors of the operator: in $D_{2d}$, $\hat{L}_z$ transforms as $A_2$ and $\hat{L}_\pm$ as $E$, so a matrix element is nonzero only if the orbital product of the two states contains the corresponding irreducible representation. Accordingly, the ISC from the singlet $^1B_2$ to the triplet $^3B_2$ is symmetry forbidden by first-order SOC, as the orbital product $B_2 \otimes B_2 = A_1$ contains neither $A_2$ (the symmetry of the orbital factor $\hat{L}_z$) nor $E$ (the symmetry of the orbital factor $\hat{L}_\pm$). By contrast, we found that the $\lvert ^1E, m_s=0\rangle$ state [Eq.~\eqref{eq:1E}], lying above the $^1B_2$ singlet bound-exciton state, could couple to $\lvert ^3B_2, m_s=\pm1\rangle$, but could not to $\lvert ^3B_2, m_s=0\rangle$ (since the orbital product $E \otimes B_2 = E$ matches the symmetry of the orbital factor $\hat{L}_\pm$ but not that of $\hat{L}_z$).

We further express the matrix elements of the two-particle spin--orbit operator in Eq.~\eqref{eq:Hso} in terms of single-particle matrix elements between the Kohn--Sham orbitals; the latter are read directly from our non-collinear DFT calculation results. The symmetry-allowed spin--orbit couplings connecting the excited singlet $\lvert ^1E\rangle$ and triplet $\lvert ^3B_2,{\pm}1\rangle$ reduce to

\begin{equation}
\label{eq:soc_plus}
\begin{split}
\langle^1E\lvert \hat{L}_+\hat{S}_-\lvert ^3B_2,m_s=+1\rangle
 = \frac{1}{2}\langle e^\downarrow \lvert \hat{L}_+\hat{S}_-\lvert b^{\uparrow}\rangle
\end{split}
\end{equation}

\begin{equation}
\label{eq:soc_minus}
\begin{split}
\langle^1E\lvert \hat{L}_-\hat{S}_+\lvert ^3B_2,m_s=-1\rangle
 = -\frac{1}{2}\langle e^\uparrow \lvert \hat{L}_-\hat{S}_+\lvert b^{\downarrow}\rangle.
\end{split}
\end{equation}

The results show that the spin--orbit coupling strength $\lambda_+$ between $\lvert ^1E, m_s=0\rangle$ and $\lvert ^3B_2, m_s=+1\rangle$ is almost equal to $\lambda_-$ between $\lvert ^1E, m_s=0\rangle$ and $\lvert ^3B_2, m_s=-1\rangle$.
The ISC channel is thus inherently spin selective: it connects $\lvert ^1E, m_s=0\rangle$ only to the $m_s=\pm1$ sublevels of the triplet (which are fed equally), leaving $m_s=0$ uncoupled.

The fate of the singlet branch is governed by the competition at the $^1E$ state between this ISC channel and the spin-conserving internal conversion (IC) toward the lower-lying singlet states, which terminates in $^1B_2$; from there only radiative recombination is available, as the direct $^1B_2\to{}^3B_2$ ISC is symmetry-forbidden. The fraction of the singlet branch shelved into the triplet is therefore
\begin{equation}
\label{eq:eta_s}
\eta_s=\frac{\Gamma_{\mathrm{ISC}}}{\Gamma_{\mathrm{ISC}}+\Gamma_{\mathrm{IC}}}\approx\frac{\Gamma_{\mathrm{ISC}}}{\Gamma_{\mathrm{IC}}},
\end{equation}
since the spin-allowed IC is expected to be by far the fastest channel out of the $^1E$ state; on this timescale all other decay channels of $^1E$, including its radiative recombination, are negligible. The singlet branch alone thus shelves only a small fraction of its population into the triplet ($\eta_s\ll1$; Sec.~\ref{sec:rates}).

Combining the two routes, the fraction of capture events that shelve population in the triplet is $\eta_T\approx3/4+\eta_s/4$, dominated by the capture statistics. An important corollary concerns spin polarization: statistical capture populates the three triplet sublevels equally, and the $m_s=\pm1$ selectivity of the weak $^1E$ ISC channel is diluted by a factor of $\sim\eta_s/3$, so the shelved population arrives essentially unpolarized.

These considerations imply a characteristic dependence of the triplet population on the excitation protocol. Above-band-gap excitation populates the triplet efficiently through the capture statistics, with a small additional contribution from the $^1E$ ISC channel of the singlet branch. Under resonant excitation of the $^1B_2$ ZPL, by contrast, the optical selection rules create the singlet exclusively, the $^1E$ state is bypassed, and the direct $^1B_2\to{}^3B_2$ ISC is symmetry-forbidden, so the triplet remains dark. This picture is directly supported by photo-ESR experiments: a strong NL51 signal was detected only for photon energies exceeding the band gap, whereas for sub-band-gap excitation only a weak signal remained, mediated by carrier capture at ionized defects\cite{mchedlidze2003physb}. The model further predicts that resonant excitation of the $^1B_2$ ZPL of the neutral defect suppresses the shelving of population into the quantum memory, which can be tested experimentally.

Once populated, the triplet is protected against decay: phosphorescence from the triplet $^3B_2$ back to the ground state $^1A_1$ is also symmetry-forbidden by first-order SOC. Vibronic coupling or symmetry lowering could partially relax this restriction and result in a weak ISC, enabling a phosphorescence process, similar to the C-center in Si\cite{udvarhelyi2022lband}. This protection is, however, bounded in temperature by the small singlet--triplet gap itself: thermal repopulation of the radiatively active $^1B_2$ singlet opens a loss channel of rate $\approx\frac{1}{3}\exp(-\Delta E_{\mathrm{ST}}/k_BT)\,\Gamma_{\mathrm{rad}}$ [orange arrow in Fig.~\ref{fig:mech}(a)]. This is the mechanism behind the observed quenching of the NL51 signal above $\approx$35 K, from whose temperature dependence $\Delta E_{\mathrm{ST}}$ was extracted\cite{mchedlidze2003physb}.

\subsubsection{Rate estimates and their reliability}
\label{sec:rates}

The quantities entering the mechanism described above are collected in Table~\ref{tab:rates}. In this subsection we describe how they were obtained and assess their reliability.

\begin{table}[htbp]
\centering
\caption{\label{tab:rates}%
Summary of the parameters and rates that govern the population, decay, and storage in the $^3B_2$ triplet state of I$_4$. $r$ is the ratio of the per-sublevel capture cross sections of the triplet and singlet manifolds.}
\begin{ruledtabular}
\begin{tabular}{ll}
Quantity & Value \\
\hline
$\Delta E_{\mathrm{ST}}$ & 1.58 meV\footnotemark[3] \\
Zero-field splitting $D$ & 22.25 MHz\footnotemark[1] ($\sim$57 MHz\footnotemark[4]) \\
$\tau_{\mathrm{rad}}$ ($\Gamma_{\mathrm{rad}}$) & 351 ns ($2.8\times10^{6}$ s$^{-1}$)\footnotemark[1] \\
$\Delta E$ ($^1E$--$^3B_2$) & $\lesssim$100 meV\footnotemark[2] \\
$|H_{\mathrm{SOC}}|$ & 0.017 meV\footnotemark[1] \\
$S$ ($^1B_2$/$^3B_2$) & 0.42\footnotemark[1] \\
$F(\Delta E)$ & 0.25 eV$^{-1}$\footnotemark[1] \\
$\Gamma_{\mathrm{ISC}}$ ($T=0$ K) & $6.9\times10^{5}$ s$^{-1}$\footnotemark[1] \\
$\Gamma_{\mathrm{IC}}$ & $\gtrsim10^{2}\,\Gamma_{\mathrm{ISC}}$\footnotemark[2] \\
$\eta_s$ & $10^{-2}$--$10^{-6}$\footnotemark[2] \\
$\eta_T$ & $\approx$3/4 (0.60--0.86 for $r=0.5$--2)\footnotemark[2] \\
Thermal loss time (2 K / 10 K) & $\sim$10 ms / $\sim$7 $\mu$s\footnotemark[2] \\
Intrinsic triplet lifetime & $\gtrsim$6 ms\footnotemark[5]
\end{tabular}
\end{ruledtabular}
\footnotetext[1]{Calculated in this work.}
\footnotetext[2]{Estimated in this work as described in the text.}
\footnotetext[3]{Experiment, Ref.~\onlinecite{mchedlidze2003physb}.}
\footnotetext[4]{Experiment, Ref.~\onlinecite{85stallinga1994epr}.}
\footnotetext[5]{From the experimental lifetime ratio of Ref.~\onlinecite{mchedlidze2003physb} and the calculated $\tau_{\mathrm{rad}}$.}
\end{table}

Evaluating the ISC rate requires the energy gap between the $^1E$ and $^3B_2$ states. A direct $\Delta$SCF calculation of the $^1E$ level is not reliable here, as the total-energy differences of such strongly delocalized states are sensitive to the constrained occupation of nearly degenerate supercell bands and to finite-size effects. We therefore estimate the gap from the Kohn--Sham eigenvalue difference of the $e$ and $b_2$ valley-orbit levels in the ground-state electron configuration, which yields $\Delta E\approx100$ meV. This estimate is justified on three grounds. First, the $^1E$ and $^1B_2$ bound-exciton states share the identical localized $a_1$ hole and differ only in the delocalized CBM-derived electron orbital, so the differences in electron--hole Coulomb, exchange, and relaxation energies largely cancel; the exchange contributions are bounded by the meV-scale singlet--triplet splittings computed above. Second, the valley--orbit splitting originates from the short-range central-cell potential of the defect, as coupling the conduction-band valleys requires large wave-vector Fourier components; the bound hole adds only a smooth long-range Coulomb potential that shifts the whole valley--orbit manifold rigidly without affecting the splitting. Third, the applied HSE06 functional is Koopmans-compliant in silicon\cite{65deak2010hse,66deak2019hybrid}, so the Kohn--Sham eigenvalue differences are meaningful quasiparticle estimates, and any residual error common to the two CBM-derived levels cancels in their difference. The $\Gamma$-point sampling introduces a finite-size error in the valley--orbit splitting (cf.\ the band-filling discussion in Sec.~II), so $\Delta E$ should be understood as an order-of-magnitude estimate.

For the proposed mechanism it is also essential that the $^1E$ state remains a bound state. Its estimated energy ($\approx$1.13 eV above the ground state) falls below the calculated band gap (1.16 eV) and the free-exciton continuum, so the complex in the $^1E$ state does not ionize into free carriers. The lowest dissociation channel is the emission of the loosely bound electron into the conduction band, leaving I$_4^+$ behind; its threshold, given by our calculated band gap, charge transition level, and ZPL energy as $(1.16-0.08)~\mathrm{eV} - 1.034~\mathrm{eV} \approx 50$ meV above the $^1B_2$ level, is nominally lower than the supercell estimate of $\Delta E$. In this comparison, however, $\Delta E\approx100$ meV must be regarded as an upper bound: the finite supercell over-concentrates the delocalized electron at the defect core and thereby enhances all overlap-driven splittings---the same finite-size effect that inflates the singlet--triplet exchange from its dilute-limit value of $\sim$1.6 meV to 13.9 meV in the 512-atom cell. The dilute-limit valley--orbit splitting is therefore expected to be substantially smaller, placing the $^1E$ level below the electron-detachment threshold as a genuinely bound intermediate state. Experimentally, the $^1E$ state should appear as a resonance above the X-line ZPL (by at most $\sim$100 meV) in photoluminescence-excitation spectroscopy, providing a direct test of these predictions.

The computed gap of $\approx$100 meV exceeds the highest phonon energy of the system (the local vibrational modes at $\approx$68 meV, Table~\ref{table2}, and the bulk spectrum terminating at $\approx$65 meV), so one-phonon emission is energetically forbidden, and the ISC is intrinsically a multiphonon process that demands the generating-function Franck--Condon treatment of Eqs.~\eqref{eq:gamma_fcdw} and \eqref{eq:gamma_Gt}. The required displacement between the $^1E$ and $^3B_2$ equilibrium geometries was approximated by that between the $^1B_2$ and $^3B_2$ geometries. This proxy is well founded: the lattice relaxation is driven by the localized $a_1$ hole, whereas the delocalized electron contributes only a vanishing charge density per atom and hence negligible forces, so the relaxed geometries of $^1E$ and $^1B_2$---which differ only in the delocalized orbital, both derived from the same conduction-band valleys---are expected to be nearly identical. The small total Huang--Rhys factor $S=0.42$ obtained for the $^1B_2$/$^3B_2$ pair, to which the same reasoning applies, confirms this picture internally.

With the calculated SOC matrix element $|H_{\mathrm{SOC}}|\approx0.017$ meV and the Franck--Condon-weighted density evaluated at this gap, $F(\Delta E)\approx0.25$ eV$^{-1}$, Eq.~\eqref{eq:gamma_isc} gives a raw value $X = |H_{\mathrm{SOC}}|^2\,F/\hbar\approx1.1\times10^5$ s$^{-1}$, corresponding to an ISC rate $\Gamma_{\mathrm{ISC}} = 2\pi X\approx6.9\times10^5$ s$^{-1}$ at 0 K\cite{goldman2015state}. Writing the rate in this form makes explicit that the $2\pi$ factor of Fermi's golden rule is retained, so that $X = \Gamma_{\mathrm{ISC}}/2\pi$ is the ordinary-frequency value and $\Gamma_{\mathrm{ISC}}$ the corresponding angular rate. The rate is suppressed by three compounding factors: the small symmetry-allowed SOC, the weak electron--phonon coupling between the singlet and triplet potential energy surfaces ($S=0.42$), and the magnitude of the gap itself, which enforces a high-order multiphonon emission whose Franck--Condon weight falls roughly exponentially with $\Delta E$ (energy-gap law). Since $F(E)$ increases with decreasing gap in this range, evaluating the rate at the upper-bound value of $\Delta E$ makes the quoted $\Gamma_{\mathrm{ISC}}$ a conservative lower estimate. The rate quoted above refers to $T=0$ K. At finite temperature, stimulated multiphonon emission enhances it through the thermal occupation factors of the accepting modes; estimating this enhancement as $(\bar{n}+1)^{p}$ with $p=\Delta E/\hbar\omega\approx6$ for dominant accepting modes near 17 meV, it amounts to less than 2\% at 35 K, reaches a factor of $\approx$2 near 100 K, and one to two orders of magnitude at room temperature. Throughout the cryogenic range in which the NL51 signal is observed ($T<35$ K)\cite{mchedlidze2003tetra}, the $T=0$ value therefore applies without correction.

The IC rate was not calculated explicitly. It obeys the same golden-rule expression as Eq.~\eqref{eq:gamma_isc}, with $H_{\mathrm{SOC}}$ replaced by the vibronic (nonadiabatic) coupling $V_{\mathrm{IC}}$ between the $E$ and $B_2$ electron states, mediated by $e$-symmetry promoting modes. If the conversion proceeds directly across the same $\approx$100 meV gap, the Franck--Condon weighted density cancels in the ratio, and $\eta_s\approx(|H_{\mathrm{SOC}}|/|V_{\mathrm{IC}}|)^2$; with $V_{\mathrm{IC}}$ expected to exceed the SOC matrix element by one to three orders of magnitude, $\eta_s\sim10^{-2}$--$10^{-6}$, and intermediate valley--orbit states would only accelerate the IC by enabling sequential one-phonon steps. The singlet branch alone thus shelves at most $\sim$1\% of its population into the triplet.

The statistical 3:1 branching at capture assumes equal capture rates into the individual singlet and triplet sublevels, an assumption that is well justified here. The capture Hamiltonian, comprising the Coulomb and electron--phonon interactions, is spin independent, and the two final manifolds share the same orbital configuration $a_1^1b_2^1$ and the same energy release to within $\Delta E_{\mathrm{ST}}\approx1.6$ meV out of $\approx$1 eV. The only interaction that distinguishes them is the exchange integral that generates $\Delta E_{\mathrm{ST}}$ itself, so the per-sublevel capture cross sections are expected to differ only at the relative level of $10^{-3}$; no spin flip is required in either channel, so no selection rule favours one manifold over the other. The branching established at capture is then preserved, because the subsequent internal conversion is spin conserving and exceeds the spin-flipping ISC by at least the factor $\eta_s^{-1}\gtrsim10^2$ estimated above---the same weak spin--orbit coupling that renders the ISC slow also protects the spin during the relaxation cascade. A departure from the statistical ratio would require a genuinely spin-dependent capture process, that is, spin-polarized free carriers or an applied magnetic field, neither of which is present under unpolarized above-band-gap excitation; conversely, any residual spin-lattice relaxation between the two manifolds at $k_BT\ll\Delta E_{\mathrm{ST}}$ would drive population toward the lower-lying triplet, so that 3/4 is a conservative estimate. The conclusion is in any case robust against moderate deviations: writing $\eta_T=3r/(3r+1)$ with $r$ the ratio of the per-sublevel capture cross sections, even $r=0.5$ or $r=2$ yields $\eta_T=60\%$ or $86\%$.

Finally, the calculated radiative lifetime of the singlet bound exciton, $\tau_{\mathrm{rad}}\approx351$ ns ($\Gamma_{\mathrm{rad}}=1/\tau_{\mathrm{rad}}\approx2.8\times10^{6}$ s$^{-1}$)---relatively long because of the delocalized nature of the excited state---is a direct prediction for the low-temperature decay time of the X-line PL. Inserted into the thermal loss rate given in Sec.~\ref{sec:mech}, it leaves $\sim$10 ms at 2 K but only $\sim$7 $\mu$s at 10 K; the reported lifetime ratio $\tau_T/\tau_S\geq1.7\times10^4$\cite{mchedlidze2003physb} together with our calculated $\tau_{\mathrm{rad}}$ implies an intrinsic triplet lifetime of $\gtrsim$6 ms, approached only below $\approx$2 K.

\subsubsection{Implications for quantum-memory operation}
\label{sec:implications}

The mechanism and the rates established above translate into the operating scheme sketched in Fig.~\ref{fig:mech}, which consists of three steps.

\textit{Loading.} Above-band-gap excitation shelves up to $\sim$75\% of the capture events in the $^3B_2$ state [step 1 in Fig.~\ref{fig:mech}(a)], whereas resonant excitation of the $^1B_2$ ZPL leaves the triplet empty; the excitation protocol thus switches the loading of the memory on and off. The singlet branch simultaneously provides the near-infrared X-line emission.

\textit{Initialization and control.} Because the shelved population arrives unpolarized, spin initialization of the memory has to rely on microwave manipulation of the spectroscopically resolved sublevels [step 2 in Fig.~\ref{fig:mech}(b)], rather than on optical spin pumping as in the nitrogen-vacancy center in diamond\cite{gali2019abinitio}. The small ZFS implies weak protection against magnetic noise but easy microwave addressing.

\textit{Storage.} The constraints imposed by the $D_{2d}$ symmetry severely suppress the ISC rate, which implies an exceptionally long spin--lattice relaxation time and positions this triplet state as an ideal candidate for a quantum memory rather than a fast-processing qubit. The electronic storage time is bounded by the thermal repopulation of the $^1B_2$ singlet: it reaches the millisecond range only below $\approx$2 K and drops to microseconds at 10 K (Table~\ref{tab:rates}), which sets the operating temperature of the electron-spin memory.

The storage capability of the triplet can be extended beyond this electronic limit by invoking proximate $^{29}$Si nuclear spins as ancilla qubits, in close analogy to the scheme proposed for the C-center in silicon\cite{udvarhelyi2022lband}. During the lifetime of the metastable triplet, the hyperfine interaction provides a two-qubit gate between the electron spin and a selected $^{29}$Si nuclear spin, so that the quantum state shelved in the triplet can be swapped onto the nuclear spin [step 3 in Fig.~\ref{fig:mech}(b)], where it is protected from electron-spin relaxation and can be stored well beyond the triplet lifetime. The hyperfine data in Table~\ref{table4} directly guide the choice of ancilla: the strongly coupled core spins ($A>5$ MHz) would dephase the electron spin and should be avoided (or removed by isotope purification), whereas weakly coupled $^{29}$Si nuclei on the fourth-neighbor shell or beyond offer resolvable but coherence-preserving couplings suitable for controlled electron--nuclear gate operations. The I$_4$ triplet may thus serve not only as an optically accessible electron-spin memory but also as an interface to a longer-lived nuclear-spin register intrinsic to the silicon host.


\section{CONCLUSIONS}
In conclusion, through first-principles calculations, we systematically investigated the electronic and optical properties of the I$_4$ complex in silicon to clarify the origin of the X-line. Our calculated ZPL, vibrational modes, and PL spectrum are consistent with previously reported experimental results, revealing that the X-line indeed originates from the bare I$_4$ complex. We also demonstrate that the large peaks beyond the ZPL arise as phonon replicas, due to strong coupling to very low energy quasi-local vibrations. According to our results, the B3 center observed in EPR, and a CTL at VBM + 0.29 eV, which correlates with it, do not seem to belong to  I$_4$. Additionally, the I$_4$ defect has a metastable triplet state lying below the singlet. Under above-band-gap excitation the triplet is populated efficiently, predominantly through the spin statistics of carrier capture, supplemented by a weak symmetry-allowed intersystem-crossing channel from the higher-lying $^1E$ singlet state. The $D_{2d}$ symmetry suppresses both intersystem crossing and phosphorescence by first-order spin--orbit coupling, implying an exceptionally long spin--lattice relaxation time. Combined with spectrally resolvable $m_s$ sublevels, microwave addressability, and remote nuclear spins to minimize decoherence, the triplet state of the I$_4$ complex constitutes a viable spin quantum memory intrinsic to silicon.

\section{ACKNOWLEDGMENT}
M.H.\ gratefully acknowledges the support of the CSC scholarship (No. 202508330433), as well as the financial support from the Zhejiang Provincial Natural Science Foundation (ZCLY24A0401), Huzhou Natural Science Foundation (2024YZ24), and NSFC (Grant No.\ 61804154), and the support from the Hefei Advanced Computing Center. P.D.\ acknowledges the support of the NSFC grant No.\ 12250710132. B.H.\ acknowledges the National Key Research and Development Program of China (Grant No.\ 2022YFA1402400), NSFC (Grant No.\ W2511008, 12088101), and NSAF (Grant No.\ U2230402). A.G.\ acknowledges the high-performance computational resources provided by KIF\"U (Governmental Agency for IT Development of Hungary) and the European Commission for the projects SPINUS (Grant No.\ 101135699) and QuSPARC (Grant No.\ 101186889). Fruitful discussions with Uwe Gerstmann about the hyperfine calculation are much appreciated.

\bibliography{I4}

@article{childs2010quantum,
  author = {A. M. Childs and W. van Dam},
  title = {Quantum algorithms for algebraic problems},
  journal = {Reviews of Modern Physics},
  volume = {82},
  pages = {1--52},
  year = {2010}
}

@article{monroe2002quantum,
  author = {C. Monroe},
  title = {Quantum information processing with atoms and photons},
  journal = {Nature},
  volume = {416},
  pages = {238--246},
  year = {2002}
}

@article{devoret2004implementing,
  author = {M. H. Devoret and J. M. Martinis},
  title = {Implementing qubits with superconducting integrated circuits},
  journal = {Quantum Information Processing},
  volume = {3},
  pages = {163--203},
  year = {2004}
}

@article{hanson2008coherent,
  author = {R. Hanson and D. D. Awschalom},
  title = {Coherent manipulation of single spins in semiconductors},
  journal = {Nature},
  volume = {453},
  pages = {1043--1049},
  year = {2008}
}

@article{bertaina2007rare,
  author = {S. Bertaina and S. Gambarelli and A. Tkachuk and I. N. Kurkin and B. Malkin and A. Stepanov and B. Barbara},
  title = {Rare-earth solid-state qubits},
  journal = {Nature Nanotechnology},
  volume = {2},
  pages = {39--42},
  year = {2007}
}

@article{toyli2013engineering,
  author = {D. M. Toyli and L. C. Bassett and B. B. Buckley and G. Calusine and D. D. Awschalom},
  title = {Engineering and quantum control of single spins in semiconductors},
  journal = {MRS Bulletin},
  volume = {38},
  pages = {139},
  year = {2013}
}

@article{balasubramanian2009ultralong,
  author = {G. Balasubramanian and P. Neumann and D. Twitchen and M. Markham and R. Kolesov and N. Mizuochi and J. Isoya and J. Achard and J. Beck and J. Tissler and V. Jacques and P. R. Hemmer and F. Jelezko and J. Wrachtrup},
  title = {Ultralong spin coherence time in isotopically engineered diamond},
  journal = {Nature Materials},
  volume = {8},
  pages = {383--387},
  year = {2009}
}

@article{gali2019abinitio,
  author = {A. Gali},
  title = {Ab initio theory of the nitrogen-vacancy center in diamond},
  journal = {Nanophotonics},
  volume = {8},
  number = {11},
  pages = {1907--1943},
  year = {2019}
}

@article{pompili2021multinode,
  author = {M. Pompili and S. L. N. Hermans and S. Baier and H. K. C. Beukers and P. C. Humphreys and R. N. Schouten and R. F. L. Vermeulen and M. J. Tiggelman and L. dos Santos Martins and B. Dirkse and S. Wehner and R. Hanson},
  title = {Realization of a multinode quantum network of remote solid-state qubits},
  journal = {Science},
  volume = {372},
  pages = {259--264},
  year = {2021}
}

@article{li2022sic,
  author = {Q. Li and J. Wang and F. Yan and J. Zhou and H. Wang and H. Liu and L. Guo and X. Zhou and A. Gali and Z. Liu and Z. Wang and K. Sun and G. Guo and J. Tang and H. Li and L. You and J. Xu and C. Li and G.-C. Guo},
  title = {Room-temperature coherent manipulation of single-spin qubits in silicon carbide with a high readout contrast},
  journal = {National Science Review},
  volume = {9},
  pages = {nwab122},
  year = {2022}
}

@article{koehl2011sic,
  author = {W. F. Koehl and B. B. Buckley and F. J. Heremans and G. Calusine and D. D. Awschalom},
  title = {Room temperature coherent control of defect spin qubits in silicon carbide},
  journal = {Nature},
  volume = {479},
  pages = {84--87},
  year = {2011}
}

@article{anderson2019control,
  author = {C. P. Anderson and A. Bourassa and K. C. Miao and G. Wolfowicz and P. J. Mintun and A. L. Crook and H. Abe and J. U. Hassan and N. T. Son and T. Ohshima and D. D. Awschalom},
  title = {Electrical and optical control of single spins integrated in scalable semiconductor devices},
  journal = {Science},
  volume = {366},
  pages = {1225--1230},
  year = {2019}
}

@article{bian2025theory,
  author = {G. Bian and G. Thiering and A. Gali},
  title = {Theory of optical spin polarization of axial divacancy and nitrogen-vacancy defects in {4H-SiC}},
  journal = {Physical Review Research},
  volume = {7},
  pages = {013320},
  year = {2025}
}

@article{durand2021nir,
  author = {A. Durand and Y. Baron and W. Redjem and T. Herzig and A. Benali and S. Pezzagna and J. Meijer and A. Yu. Kuznetsov and J.-M. G{\'e}rard and I. Robert-Philip and M. Abbarchi and V. Jacques and G. Cassabois and A. Dr{\'e}au},
  title = {Broad diversity of near-infrared single-photon emitters in silicon},
  journal = {Physical Review Letters},
  volume = {126},
  pages = {083602},
  year = {2021}
}

@article{khoury2022silicon,
  author = {M. Khoury and M. Abbarchi},
  title = {A bright future for silicon in quantum technologies},
  journal = {Journal of Applied Physics},
  volume = {131},
  pages = {200901},
  year = {2022}
}

@article{zhang2020platforms,
  author = {G. Zhang and Y. Cheng and J.-P. Chou and A. Gali},
  title = {Material platforms for defect qubits and single-photon emitters},
  journal = {Applied Physics Reviews},
  volume = {7},
  pages = {031308},
  year = {2020}
}

@article{song1990carbon,
  author = {L. W. Song and X. D. Zhan and B. W. Benson and G. D. Watkins},
  title = {Bistable interstitial-carbon-substitutional-carbon pair in silicon},
  journal = {Physical Review B},
  volume = {42},
  pages = {5765},
  year = {1990}
}

@article{davies1989optical,
  author = {G. Davies},
  title = {The optical properties of luminescence centres in silicon},
  journal = {Physics Reports},
  volume = {176},
  pages = {83},
  year = {1989}
}

@article{redjem2020telecom,
  author = {W. Redjem and A. Durand and T. Herzig and A. Benali and S. Pezzagna and J. Meijer and A. Y. Kuznetsov and H. S. Nguyen and S. Cueff and J.-M. G{\'e}rard and I. Robert-Philip and B. Gil and D. Caliste and P. Pochet and M. Abbarchi and V. Jacques and A. Dr{\'e}au and G. Cassabois},
  title = {Single artificial atoms in silicon emitting at telecom wavelengths},
  journal = {Nature Electronics},
  volume = {3},
  pages = {738},
  year = {2020}
}

@article{hollenbach2020telecom,
  author = {M. Hollenbach and Y. Berenc{\'e}n and U. Kentsch and M. Helm and G. V. Astakhov},
  title = {Engineering telecom single-photon emitters in silicon for scalable quantum photonics},
  journal = {Optics Express},
  volume = {28},
  pages = {26111},
  year = {2020}
}

@article{baron2022gcenters,
  author = {Y. Baron and A. Durand and T. Herzig and M. Khoury and S. Pezzagna and J. Meijer and I. Robert-Philip and M. Abbarchi and J.-M. Hartmann and S. Reboh and J.-M. G{\'e}rard and V. Jacques and G. Cassabois and A. Dr{\'e}au},
  title = {Single {G} centres in silicon fabricated by co-implantation with carbon and proton},
  journal = {Applied Physics Letters},
  volume = {121},
  pages = {084003},
  year = {2022}
}

@article{hollenbach2022wafer,
  author = {M. Hollenbach and N. Klingner and N. S. Jagtap and L. Bischoff and C. Fowley and U. Kentsch and G. Hlawacek and A. Erbe and N. V. Abrosimov and M. Helm and Y. Berenc{\'e}n and G. V. Astakhov},
  title = {Wafer-scale nanofabrication of telecom single-photon emitters in silicon},
  journal = {Nature Communications},
  volume = {13},
  pages = {7683},
  year = {2022}
}

@article{redjem2023defect,
  author = {W. Redjem and A. J. Amsellem and F. I. Allen and G. Benndorf and J. Bin and S. Bulanov and E. Esarey and L. C. Feldman and J. Ferrer Fernandez and J. Garcia Lopez and L. Geulig and C. R. Geddes and H. Hijazi and Q. Ji and V. Ivanov and B. Kant{\'e} and A. Gonsalves and J. Meijer and K. Nakamura and A. Persaud and I. Pong and L. Obst-Huebl and P. A. Seidl and J. Simoni and C. Schroeder and S. Steinke and L. Z. Tan and R. Wunderlich and B. Wynne and T. Schenkel},
  title = {Defect engineering of silicon with ion pulses from laser acceleration},
  journal = {Communications Materials},
  volume = {4},
  pages = {22},
  year = {2023}
}

@article{prabhu2023artificial,
  author = {M. Prabhu and C. Errando-Herranz and L. De Santis and I. Christen and C. Chen and C. Gerlach and D. Englund},
  title = {Individually addressable and spectrally programmable artificial atoms in silicon photonics},
  journal = {Nature Communications},
  volume = {14},
  pages = {2380},
  year = {2023}
}

@article{lefaucher2023cavity,
  author = {B. Lefaucher and J.-B. Jager and V. Calvo and A. Durand and Y. Baron and V. Jacques and I. Robert-Philip and G. Cassabois and T. Herzig and J. Meijer and S. Pezzagna and M. Khoury and M. Abbarchi and A. Dr{\'e}au and J.-M. G{\'e}rard},
  title = {Cavity-enhanced zero-phonon emission from an ensemble of {G} centers in a silicon-on-insulator microring},
  journal = {Applied Physics Letters},
  volume = {122},
  pages = {061109},
  year = {2023}
}

@article{redjem2023naturecomm,
  author = {W. Redjem and Y. Zhiyenbayev and W. Qarony and V. Ivanov and C. Papapanos and W. Liu and K. Jhuria and Z. Y. Al Balushi and S. Dhuey and A. Schwartzberg and L. Z. Tan and T. Schenkel and B. Kant{\'e}},
  title = {All-silicon quantum light source by embedding an atomic emissive center in a nanophotonic cavity},
  journal = {Nature Communications},
  volume = {14},
  number = {1},
  pages = {3321},
  year = {2023}
}

@article{durand2024genuine,
  author = {A. Durand and Y. Baron and F. Cache and T. Herzig and M. Khoury and S. Pezzagna and J. Meijer and J.-M. Hartmann and S. Reboh and M. Abbarchi and I. Robert-Philip and J.-M. G{\'e}rard and V. Jacques and G. Cassabois and A. Dr{\'e}au},
  title = {Genuine and faux single {G} centers in carbon-implanted silicon},
  journal = {Physical Review B},
  volume = {110},
  pages = {L020102},
  year = {2024}
}

@article{udvarhelyi2021telecom,
  author = {P. Udvarhelyi and B. Somogyi and G. Thiering and A. Gali},
  title = {Identification of a telecom wavelength single photon emitter in silicon},
  journal = {Physical Review Letters},
  volume = {127},
  pages = {196402},
  year = {2021}
}

@article{leary1997carbon,
  author = {P. Leary and R. Jones and S. {\"O}berg and V. J. B. Torres},
  title = {Dynamic properties of interstitial carbon and carbon-carbon pair defects in silicon},
  journal = {Physical Review B},
  volume = {55},
  pages = {2188},
  year = {1997}
}

@article{wang2014carbon,
  author = {H. Wang and A. Chroneos and C. A. Londos and E. N. Sgourou and U. Schwingenschl{\"o}gl},
  title = {Carbon related defects in irradiated silicon revisited},
  journal = {Scientific Reports},
  volume = {4},
  pages = {4909},
  year = {2014}
}

@article{deak2024telecom,
  author = {P. De{\'a}k and S. Li and A. Gali},
  title = {Quantum bit with telecom wavelength emission from a simple defect in {Si}},
  journal = {Communications Physics},
  volume = {7},
  pages = {337},
  year = {2024}
}

@article{aberl2024epitaxial,
  author = {J. Aberl and E. Prado Navarrete and M. Karaman and D. Haya Enriquez and C. Wilflingseder and A. Salomon and D. Primetzhofer and M. A. Schubert and G. Capellini and T. Fromherz and P. De{\'a}k and P. Udvarhelyi and S. Li and {\'A}. Gali and M. Brehm},
  title = {All-epitaxial self-assembly of silicon color centers confined within sub-nanometer thin layers},
  journal = {Advanced Materials},
  volume = {36},
  pages = {2408424},
  year = {2024}
}

@article{salomon2025led,
  author = {A. Salomon and J. Aberl and E. Prado Navarrete and M. Karaman and O. E. Lang and D. Primetzhofer and P. De{\'a}k and {\'A}. Gali and T. Fromherz and M. Brehm},
  title = {Telecom light-emitting diodes based on nanoconfined self-assembled silicon-based color centers},
  journal = {ACS Photonics},
  volume = {12},
  pages = {2364},
  year = {2025}
}

@article{udvarhelyi2022lband,
  author = {P. Udvarhelyi and A. Pershin and P. De{\'a}k and A. Gali},
  title = {An {L}-band emitter with quantum memory in silicon},
  journal = {npj Computational Materials},
  volume = {8},
  pages = {262},
  year = {2022}
}

@article{hayama2004isotope,
  author = {S. Hayama and G. Davies and J. Coutinho and R. Jones and K. M. Itoh},
  title = {Lattice isotope effects on optical transitions in silicon},
  journal = {Physical Review B},
  volume = {70},
  pages = {035202},
  year = {2004}
}

@article{wen2025arxiv,
  author = {S. Wen and G. Pieplow and J. Yang and K. Jamshidi and M. Helm and J.-W. Luo and T. Schr{\"o}der and S. Zhou and Y. Berenc{\'e}n},
  title = {Optical spin readout of a silicon color center in the telecom {L}-band},
  journal = {arXiv preprint arXiv:2502.07632},
  year = {2025}
}

@article{ivanov2022localization,
  author = {V. Ivanov and J. Simoni and Y. Lee and W. Liu and K. Jhuria and W. Redjem and Y. Zhiyenbayev and C. Papapanos and W. Qarony and B. Kant{\'e} and A. Persaud and T. Schenkel and L. Z. Tan},
  title = {Effect of localization on photoluminescence and zero-field splitting of silicon color centers},
  journal = {Physical Review B},
  volume = {106},
  pages = {134107},
  year = {2022}
}

@article{bergeron2020interface,
  author = {L. Bergeron and C. Chartrand and A. T. K. Kurkjian and K. J. Morse and H. Riemann and N. V. Abrosimov and P. Becker and H.-J. Pohl and M. L. W. Thewalt and S. Simmons},
  title = {Silicon-integrated telecommunications photon-spin interface},
  journal = {PRX Quantum},
  volume = {1},
  pages = {020301},
  year = {2020}
}

@article{higginbottom2022single,
  author = {D. B. Higginbottom and A. T. K. Kurkjian and C. Chartrand and M. Kazemi and N. A. Brunelle and E. R. MacQuarrie and J. R. Klein and N. R. Lee-Hone and J. Stacho and M. Ruether and C. Bowness and L. Bergeron and A. DeAbreu and S. R. Harrigan and J. Kanaganayagam and D. W. Marsden and T. S. Richards and L. A. Stott and S. Roorda and K. J. Morse and M. L. W. Thewalt and S. Simmons},
  title = {Optical observation of single spins in silicon},
  journal = {Nature},
  volume = {607},
  pages = {266},
  year = {2022}
}

@article{safonov1996carbon,
  author = {A. N. Safonov and E. C. Lightowlers and G. Davies and P. Leary and R. Jones and S. {\"O}berg},
  title = {Interstitial-carbon hydrogen interaction in silicon},
  journal = {Physical Review Letters},
  volume = {77},
  pages = {4812},
  year = {1996}
}

@article{dhaliah2022tcenter,
  author = {D. Dhaliah and Y. Xiong and A. Sipahigil and S. M. Griffin and G. Hautier},
  title = {First-principles study of the {T} center in silicon},
  journal = {Physical Review Materials},
  volume = {6},
  pages = {L053201},
  year = {2022}
}

@article{xiong2024discovery,
  author = {Y. Xiong and J. Zheng and S. McBride and X. Zhang and S. M. Griffin and G. Hautier},
  title = {Computationally driven discovery of {T} center-like quantum defects in silicon},
  journal = {Journal of the American Chemical Society},
  volume = {146},
  pages = {30046},
  year = {2024}
}

@article{buckley2020wcenters,
  author = {S. M. Buckley and A. Tait and G. Moody and B. Primavera and S. Olson and J. Herman and K. L. Silverman and S. Papa Rao and S. W. Nam and R. P. Mirin and J. M. Shainline},
  title = {Optimization of photoluminescence from {W} centers in silicon-on-insulator},
  journal = {Optics Express},
  volume = {28},
  pages = {16057--16072},
  year = {2020}
}

@article{bao2007led,
  author = {J. Bao and M. Tabbal and T. Kim and S. Charnvanichborikarn and J. S. Williams and M. J. Aziz and F. Capasso},
  title = {Point defect engineered {Si} sub-bandgap light-emitting diode},
  journal = {Optics Express},
  volume = {15},
  number = {11},
  pages = {6727--6733},
  year = {2007}
}

@article{komza2024indistinguishable,
  author = {L. Komza and P. Samutpraphoot and M. Odeh and Y. L. Tang and M. Mathew and J. Chang and H. Song and M.-K. Kim and Y. Xiong and G. Hautier and A. Sipahigil},
  title = {Indistinguishable photons from an artificial atom in silicon photonics},
  journal = {Nature Communications},
  volume = {15},
  pages = {6920},
  year = {2024}
}

@article{chartrand2018linewidth,
  author = {C. Chartrand and L. Bergeron and K. J. Morse and H. Riemann and N. V. Abrosimov and P. Becker and H.-J. Pohl and S. Simmons and M. L. W. Thewalt},
  title = {Highly enriched $^{28}${Si} reveals remarkable optical linewidths and fine structure for well-known damage centers},
  journal = {Physical Review B},
  volume = {98},
  pages = {195201},
  year = {2018}
}

@article{tan2003implantation,
  author = {J. Tan and G. Davies and S. Hayama and R. Harding and J. Wong-Leung},
  title = {Ion implantation effects in silicon with high carbon content characterized by photoluminescence},
  journal = {Physica B},
  volume = {340--342},
  pages = {714},
  year = {2003}
}

@article{lee2008interstitial,
  author = {S. Lee and G. S. Hwang},
  title = {Structure and stability of small compact self-interstitial clusters in crystalline silicon},
  journal = {Physical Review B},
  volume = {77},
  pages = {085210},
  year = {2008}
}

@article{baron2022wcenters,
  author = {Y. Baron and A. Durand and P. Udvarhelyi and T. Herzig and M. Khoury and S. Pezzagna and J. Meijer and I. Robert-Philip and M. Abbarchi and J.-M. Hartmann and V. Mazzocchi and J.-M. G{\'e}rard and A. Gali and V. Jacques and G. Cassabois and A. Dr{\'e}au},
  title = {Detection of single {W}-centers in silicon},
  journal = {ACS Photonics},
  volume = {9},
  pages = {2337},
  year = {2022}
}

@inproceedings{tkachev1977radiative,
  author = {V. O. Tkachev and A. V. Mudryi},
  title = {Radiative recombination centres in silicon irradiated by fast neutrons and ions},
  booktitle = {Institute of Physics Conference Series},
  volume = {31},
  pages = {231},
  year = {1977}
}

@article{awadelkarim1990photoluminescence,
  author = {O. O. Awadelkarim and A. Henry and B. Monemar and J. L. Lindstr{\"o}m and Y. Zhang and J. W. Corbett},
  title = {Photoluminescence study of radiative channels in ion-implanted silicon},
  journal = {Physical Review B},
  volume = {42},
  pages = {5635},
  year = {1990}
}

@article{52hayama2004implantation,
  author = {S. Hayama and G. Davies and K. M. Itoh},
  title = {Photoluminescence studies of implantation damage centers in $^{30}${Si}},
  journal = {Journal of Applied Physics},
  volume = {96},
  number = {3},
  pages = {1754--1756},
  year = {2004},
  doi = {10.1063/1.1767965}
}

@article{arai1997interstitial,
  author = {N. Arai and S. Takeda and M. Kohyama},
  title = {Self-interstitial clustering in crystalline silicon},
  journal = {Physical Review Letters},
  volume = {78},
  pages = {4265},
  year = {1997}
}

@article{54carvalho2005dft,
  author = {A. Carvalho and R. Jones and J. Coutinho and P. R. Briddon},
  title = {Density-functional study of small interstitial clusters in {Si}: {C}omparison with experiments},
  journal = {Physical Review B},
  volume = {72},
  pages = {155208},
  year = {2005}
}

@article{santos2016origin,
  author = {I. Santos and M. Aboy and P. L{\'o}pez and L. A. Marqu{\'e}s and L. Pelaz},
  title = {Insights on the atomistic origin of {X} and {W} photoluminescence lines in c-{Si} from ab initio simulations},
  journal = {Journal of Physics D: Applied Physics},
  volume = {49},
  pages = {075109},
  year = {2016}
}

@article{pierreux2005esr,
  author = {D. Pierreux and A. Stesmans},
  title = {Electron spin resonance study of paramagnetic centers in neutron-irradiated heat-treated silicon},
  journal = {Physical Review B},
  volume = {71},
  pages = {115204},
  year = {2005}
}

@article{brower1976epr,
  author = {K. L. Brower},
  title = {{EPR} of a (001) {Si} interstitial complex in irradiated silicon},
  journal = {Physical Review B},
  volume = {14},
  pages = {872},
  year = {1976}
}

@article{mchedlidze2003tetra,
  author = {T. Mchedlidze and I. Yonenaga and M. Suezawa},
  title = {{ESR} signature of tetra-interstitial defect in silicon},
  journal = {Materials Science in Semiconductor Processing},
  volume = {6},
  pages = {263--266},
  year = {2003}
}

@article{mchedlidze2003physb,
  author = {T. Mchedlidze and M. Suezawa},
  title = {Properties of tetra-interstitial agglomerate in silicon: an {ESR} study},
  journal = {Physica B},
  volume = {340-342},
  pages = {682--686},
  year = {2003}
}

@article{59newman1975optical,
  author = {R. C. Newman and D. H. J. Totterdell},
  title = {An optical study of defects in silicon irradiated with fast neutrons},
  journal = {Journal of Physics C: Solid State Physics},
  volume = {8},
  pages = {3944},
  year = {1975}
}

@article{mukashev1982defects,
  author = {B. N. Mukashev and A. V. Spitsyn and N. Fukuoka and H. Saito},
  title = {Defects in carbon implanted silicon},
  journal = {Japanese Journal of Applied Physics},
  volume = {21},
  pages = {399},
  year = {1982}
}

@article{coomer2001tetra,
  author = {B. J. Coomer and J. P. Goss and R. Jones and S. {\"O}berg and P. R. Briddon},
  title = {Identification of the tetra-interstitial in silicon},
  journal = {Journal of Physics: Condensed Matter},
  volume = {13},
  pages = {L1},
  year = {2001}
}

@article{kiyoi2022annealing,
  author = {A. Kiyoi and N. Kawabata and K. Nakamura and Y. Fujiwara},
  title = {Annealing-temperature-dependent evolution of hydrogen-related donor and its strong correlation with {X}-photoluminescence center in proton-irradiated silicon},
  journal = {Journal of Applied Physics},
  volume = {131},
  pages = {125702},
  year = {2022}
}

@article{myers1992hydrogen,
  author = {S. M. Myers and M. I. Baskes and H. K. Birnbaum and J. W. Corbett and G. G. DeLeo and S. K. Estreicher and E. E. Haller and P. Jena and N. M. Johnson and R. Kirchheim and S. J. Pearton and M. J. Stavola},
  title = {Hydrogen interactions with defects in crystalline solids},
  journal = {Reviews of Modern Physics},
  volume = {64},
  pages = {559},
  year = {1992}
}

@article{kiyoi2023hydrogen,
  author = {A. Kiyoi and T. Umeda},
  title = {How does hydrogen transform into shallow donors in silicon?},
  journal = {Physical Review B},
  volume = {108},
  pages = {235201},
  year = {2023}
}

@article{65deak2010hse,
  author = {P. De{\'a}k and B. Aradi and T. Frauenheim and E. Janz{\'e}n and A. Gali},
  title = {Accurate defect levels obtained from the {HSE06} range-separated hybrid functional},
  journal = {Physical Review B},
  volume = {81},
  pages = {153203},
  year = {2010}
}

@article{66deak2019hybrid,
  author = {P. De{\'a}k and M. Lorke and B. Aradi and T. Frauenheim},
  title = {Optimized hybrid functionals for defect calculations in semiconductors},
  journal = {Journal of Applied Physics},
  volume = {126},
  pages = {130901},
  year = {2019}
}

@article{67kresse1994md,
  author = {G. Kresse and J. Hafner},
  title = {Ab initio molecular-dynamics simulation of the liquid-metal--amorphous-semiconductor transition in germanium},
  journal = {Physical Review B},
  volume = {49},
  pages = {14251},
  year = {1994}
}

@article{68kresse1996vasp,
  author = {G. Kresse and J. Furthm{\"u}ller},
  title = {Efficient iterative schemes for ab initio total-energy calculations using a plane-wave basis set},
  journal = {Physical Review B},
  volume = {54},
  pages = {11169},
  year = {1996}
}

@article{69kresse1999paw,
  author = {G. Kresse and D. Joubert},
  title = {From ultrasoft pseudopotentials to the projector augmented-wave method},
  journal = {Physical Review B},
  volume = {59},
  pages = {1758},
  year = {1999}
}

@article{70chagas2021correction,
  author = {M. Chagas de Silva and M. Lorke and B. Aradi and M. Farzalipour Tabriz and T. Frauenheim and A. Rubio and D. Rocca and P. De{\'a}k},
  title = {Self-consistent potential correction for charged periodic systems},
  journal = {Physical Review Letters},
  volume = {126},
  pages = {076401},
  year = {2021}
}

@article{71gali2023review,
  author = {A. Gali},
  title = {Recent advances in the ab initio theory of solid-state defect qubits},
  journal = {Nanophotonics},
  volume = {12},
  pages = {359},
  year = {2023}
}

@article{72mackoit2019carbon,
  author = {M. Mackoit-Sinkevi{\v{c}}ien{\.{e}} and M. Maciaszek and C. G. Van de Walle and A. Alkauskas},
  title = {Carbon dimer defect as a source of the 4.1 eV luminescence in hexagonal boron nitride},
  journal = {Applied Physics Letters},
  volume = {115},
  pages = {212101},
  year = {2019}
}

@article{73lany2008assessment,
  author = {S. Lany and A. Zunger},
  title = {Assessment of correction methods for the band-gap problem and for finite-size effects in supercell defect calculations},
  journal = {Physical Review B},
  volume = {78},
  pages = {235104},
  year = {2008}
}

@article{74massidda1993hf,
  author = {S. Massidda and M. Posternak and A. Baldereschi},
  title = {{H}artree-{F}ock {LAPW} approach to the electronic properties of periodic systems},
  journal = {Physical Review B},
  volume = {48},
  pages = {5058},
  year = {1993}
}

@article{75broqvist2009erratum,
  author = {P. Broqvist and A. Alkauskas and A. Pasquarello},
  title = {Erratum: {H}ybrid-functional calculations with plane-wave basis sets},
  journal = {Physical Review B},
  volume = {80},
  pages = {085114},
  year = {2009}
}

@article{76ho2018ga2o3,
  author = {Q. D. Ho and T. Frauenheim and P. De{\'a}k},
  title = {Origin of photoluminescence in $\beta$-{Ga$_2$O$_3$}},
  journal = {Physical Review B},
  volume = {97},
  pages = {115163},
  year = {2018}
}

@article{77alkauskas2014lineshape,
  author = {A. Alkauskas and B. B. Buckley and D. D. Awschalom and C. G. Van de Walle},
  title = {First-principles theory of the luminescence lineshape for the triplet transition in diamond {NV} centres},
  journal = {New Journal of Physics},
  volume = {16},
  pages = {073026},
  year = {2014}
}

@article{78perdew1996gga,
  author = {J. P. Perdew and K. Burke and M. Ernzerhof},
  title = {Generalized gradient approximation made simple},
  journal = {Physical Review Letters},
  volume = {77},
  pages = {3865},
  year = {1996}
}

@article{79li2022ws2,
  author = {S. Li and G. Thiering and P. Udvarhelyi and V. Iv{\'a}dy and A. Gali},
  title = {Carbon defect qubit in two-dimensional {WS$_2$}},
  journal = {Nature Communications},
  volume = {13},
  pages = {1210},
  year = {2022}
}

@article{80green2008silicon,
  author = {M. A. Green},
  title = {Self-consistent optical parameters of intrinsic silicon at 300 {K} including temperature coefficients},
  journal = {Solar Energy Materials and Solar Cells},
  volume = {92},
  pages = {1305},
  year = {2008}
}

@article{81bodrog2013zfs,
  author = {Z. Bodrog and A. Gali},
  title = {The spin--spin zero-field splitting tensor in the projector-augmented-wave method},
  journal = {Journal of Physics: Condensed Matter},
  volume = {26},
  pages = {015305},
  year = {2013}
}

@article{82szasz2013hyperfine,
  author = {K. Sz{\'a}sz and T. Hornos and M. Marsman and A. Gali},
  title = {Hyperfine coupling of point defects in semiconductors by hybrid density functional calculations},
  journal = {Physical Review B},
  volume = {88},
  pages = {075202},
  year = {2013}
}

@article{83xiong2025dscf,
  author = {Y. Xiong and G. Hautier},
  title = {Delta {SCF} in {VASP} for excited-state defect computations: tips and pitfalls},
  journal = {arXiv preprint arXiv:2505.04748},
  year = {2025}
}

@article{84tawfik2022pyphotonics,
  author = {S. A. Tawfik and S. P. Russo},
  title = {Py{P}hotonics: {A} {P}ython package for the evaluation of luminescence properties of defects},
  journal = {Computer Physics Communications},
  volume = {273},
  pages = {108222},
  year = {2022}
}

@article{85stallinga1994epr,
  author = {P. Stallinga and T. Gregorkiewicz and C. A. J. Ammerlaan and Yu. V. Gorelkinskii},
  title = {Electron paramagnetic resonance study of the {NL51} spectrum in hydrogen implanted silicon},
  journal = {Solid State Communications},
  volume = {90},
  pages = {401--404},
  year = {1994}
}

@article{goldman2015state,
  author = {M. L. Goldman and M. W. Doherty and A. Sipahigil and N. Y. Yao and S. D. Bennett and N. B. Manson and A. Kubanek and M. D. Lukin},
  title = {State-selective intersystem crossing in nitrogen-vacancy centers},
  journal = {Physical Review B},
  volume = {91},
  number = {16},
  pages = {165201},
  year = {2015}
}

@article{elsayed1963soc,
  author = {M. A. El-Sayed},
  title = {Spin--Orbit Coupling and the Radiationless Processes in Nitrogen Heterocyclics},
  journal = {The Journal of Chemical Physics},
  volume = {38},
  number = {12},
  pages = {2834--2838},
  year = {1963},
  doi = {10.1063/1.1733610}
}

\end{document}